\documentstyle[12pt,aaspp4]{article}
\input psfig.sty
\def\rhii{\mbox {${\rm R}_{\rm HII}$} }

\def\h2{\mbox { ${\rm H}_2$} }
\def\c2{\mbox {\rm cm}^{-2}}

\def\mug{\mbox {$\mu_{gas}$} }
\def\mucrit{\mbox {$\mu_{crit}$} }
\def\alp{\mbox {$\alpha$} }
\def\co{\mbox {$^{12}{\rm CO}~(J=1-0)$} }
\def\IZw18{I~Zw~18}
\def\m82{M82}

\def\deg{\mbox {$^{\circ}$}}
\def\msun{\mbox {${\rm ~M_\odot}$}}

\def\msunyr{\mbox {$~{\rm M_\odot}$~yr$^{-1}$}}
\def\msunyrk2{\mbox {$~{\rm M_\odot}$~yr$^{-1}$~kpc$^{-2}$}}
\def\msunpc2{\mbox {${\rm ~M_\odot ~pc}^{-2}$}}

\def\Ha{\mbox {H$\alpha$~}}

\def\o3hb{[OIII]$\lambda5007$~/~H$\beta$~}
\def\O1ha{[OI]$\lambda6300$~/~H$\alpha$~}

\def\s2ha{[SII]$\lambda\lambda6717,31$~/~H$\alpha$~}
\def\2z2{HeII~$\lambda4686$~}
\def\z7{[NII]~$\lambda6583$ }
\def\N2{[NII]~$\lambda6583$~/~H$\alpha$~}
\def\16z2{[SII]~$\lambda\lambda6717, 6731$ }

\def\n{NGC~}
\def\asec{\ifmmode {'' }\else $''~$\fi}  
\def\amin{\ifmmode {' }\else $'~$\fi}    
\def\arcsper{\ifmmode \rlap.{'' }\else $\rlap{.}'' $\fi} 
\def\arcmper{\ifmmode \rlap.{' }\else $\rlap{.}' $\fi} 
\def\sles{\lower2pt\hbox{$\buildrel {\scriptstyle <}
   \over {\scriptstyle\sim}$}} 
\def\sgreat{\lower2pt\hbox{$\buildrel {\scriptstyle >}
    \over {\scriptstyle\sim}$}} 
\def\kms{\mbox {~km~s$^{-1}$} }

\def\cm3{~cm$^{-3}$}

\def\fig{{Figure}}

\def\et{{\rm et\thinspace al.}\ }   

\def\apj{ApJ}
\def\apjs{ApJS}

\def\aj{AJ}
\def\mn{MNRAS}

\def\aa{A\&A}
\def\aasup{A\&AS}

\def\annrev{ARA\&R}

\begin{document}

\title{Star Formation Thresholds in Galactic Disks\altaffilmark{1}}

\author{Crystal L. Martin}
\affil{Astronomy Department, California Institute of Technology,
MC 105-24, Pasadena, CA 91125}

\and

\author{Robert C. Kennicutt, Jr.}
\affil{Steward Observatory, University of Arizona, Tucson, AZ 85721}

\altaffiltext{1}{Data obtained in part at Kitt Peak National Observatory,
operated by AURA under contract to the National Science Foundation.}


\begin{abstract}
We report the first results of a detailed study of the star formation 
law in a sample of 32 nearby spiral galaxies with well-measured rotation 
curves, HI and H$_2$ (as traced by CO) surface density profiles, and new 
\Ha CCD photometry.  In this paper we present an atlas of \Ha images and
radial surface brightness profiles and describe a surface-density threshold 
in the star formation law.  Prominent breaks in the \Ha surface-brightness 
profiles are identified in nearly all of the actively star-forming disks, 
confirming previous claims of star formation thresholds based on lower 
quality data.  We measure the ratio of the gas density to the critical
density for local gravitational stability at the threshold radii.
The outer threshold radii observed in Sab--Sdm galaxies are in 
general agreement with those expected from the Toomre~Q stability
criterion, confirming earlier work,  but with a significant variation
that appears to be weakly correlated with galaxy type. Such a trend could 
plausibly reflect variations in the relative contribution of 
the stellar disk to the instability of the gas disk across this range
of galaxy types.  Among disks with sub-critical gas surface densities,
and outside the threshold radius in star-forming disks, the number of
isolated HII regions increases as the gas surface density approaches
the critical density. At the thresholds, the gas surface densities span a 
wide range; and the atomic/molecular gas fraction is highest in the disks
having the lowest total gas surface density. The simple Toomre condition 
fails to account for the active star formation in the inner disks of 
low-mass spirals such as NGC~2403 and M33. An alternative stability criterion 
based on the shear in the disk  provides a better description of these 
disks but is a less accurate indicator of the outer edges of star-forming 
disks than the Toomre criterion.  These results strongly support the view that 
the formation of gravitationally bound interstellar clouds regulates the 
onset of widespread star formation -- at least in the outer regions
of galactic disks.  
\end{abstract}


\section{Introduction}
The evolution of galaxies is strongly influenced by how quickly 
gas is consumed by stars.  Most simulations assume the star
formation rate (SFR) scales as a power law of the gas density as
originally proposed by Schmidt (1959) or adopt the surface density 
parameterization introduced by Kennicutt (1989).  A global correlation 
between SFR and gas surface density ($\Sigma_{SFR} \propto \mu^{1.4}_{gas}$)
is obtained empirically when both quantities are normalized by the area of 
the stellar disk.  Since galaxies typically have gaseous disks that are 
significantly larger than their stellar disks (e.g. Warmels 1988a; Warmels  
1988b; Gallagher \& Hunter 1984), the gas in the outer disk is not included 
in the surface density measurement (Kennicutt 1998).  The empirical 
star formation recipe implicitly includes a surface density 
threshold for star formation, and this should be incorporated in the
prescriptions used in numerical simulations.


The idea that gravitational instability might determine the critical gas
density for star formation was introduced by Spitzer (1968) and Quirk (1972) 
soon after local stability criteria for differentially rotating disks were
developed (Toomre 1964; Goldreich \& Lynden-Bell 1965). A thin, gas disk is 
unstable to axisymmetric disturbances where the Toomre Q~parameter,
	\begin{equation}
	 Q(R) \equiv \frac{\sigma \kappa }{\pi G \mu}, 
	\end{equation}
is less than unity.
The epicyclic frequency,  $\kappa$,  velocity dispersion, $\sigma$, and
surface density, $\mu$, refer to the gas disk at galactocentric radius R.
Widespread star formation is expected where the gas surface density exceeds
the critical surface density defined as
	\begin{equation}
	\mucrit = \alpha_Q \frac{\sigma \kappa}{\pi G},
	\label{eqn:mucrit}
	\end{equation}
The parameter $\alpha_Q$ is fitted to the threshold values of
the radially varying quantity
	\begin{equation}
	\alp(R) = \mug(R) / \mucrit(R)
	\end{equation} 
and makes allowances for deviations from the idealized thin 
disk model such as  finite scaleheight or the presence of a stellar disk.
The velocity dispersion is predicted to remain roughly constant 
in self-regulated regions of disks (Silk 1997) and may have a lower
bound set by the dissipation of MHD-driven turbulence in the outer disk 
(Sellwood \& Balbus 1999). 
If the gas velocity dispersion does not vary much with radius across a 
spiral galaxy, then the critical surface density falls as roughly $R^{-1}$. 
This gradient is shallower than the decline in the total gas surface density 
with radius, so the critical density will exceed the gas surface density
at some threshold radius.  Gravitational stability therefore provides an 
appealing explanation for the rather sharp edges of stellar disks 
(van der Kruit \& Searle 1981a,b; 1982a,b).

Kennicutt (1989) directly tested this hypothesis using the radial 
distribution of HII regions to trace the SFR.  He adopted a constant
gas velocity dispersion of 6\kms and found the Schmidt law  
broke down at radii where the gas surface density was less than 
0.63\mucrit (Kennicutt 1989\footnote{
	K89 reported a value of $\alpha_Q = 0.67$ but
	used a constant of 3.36	rather than $\pi$ in eq. (1).  
	}
).  The gas surface density in low surface brightness galaxies was 
subsequently shown to be below the critical density for gravitational
instability, and gravitational thresholds were used to explain the low 
level of star formation activity in these gas-rich galaxies
(van der Hulst \et 1993; van Zee \et 1997; van Zee \et 1996). More recently, 
Ferguson \et (1998) have questioned the validity of a constant gas 
velocity dispersion.  Since measurements of the HI velocity dispersion
demand both high angular resolution and high brightness sensitivity --
requirements not met by most observations, the most feasible way forward
currently is a comparison of measured threshold radii to the
thresholds predicted using constant velocity dispersion 
in a large sample of galaxies.

%

Hunter \et (1998) have recently questioned the utility
of the Toomre criterion for describing star-formation thresholds.
They measure a mean value of $\alpha_Q$ in irregular galaxies that 
is a factor of
two lower than that found by K89 for spiral galaxies.  This result implies
that the gas in the irregular galaxies is less stable than
the gas in spiral galaxies. In addition to the absence of spiral density
waves in irregular galaxies, these disks have less rotational shear
on average. (i.e. Their rotation curves are closer to solid body.)
If the local shear rate, rather than the Coriolis force (essentially 
$\kappa$), best describes the destruction rate of giant clouds,
then the stability criterion would need to be modified
(Elmegreen 1987; Elmegreen 1993; Elmegreen 1991). For example,
the local shear rate is described by the Oort A constant, 
	\begin{equation}
	A = -0.5 R \frac{d\Omega}{dR}.
	\end{equation}  
Following Elmegreen (1993) and Hunter, Elmegreen, \& Baker 
(1998, hereafter HEB98)
shearing perturbations grow at the rate $\pi G \mu/\sigma$,
and the critical column density for significant growth in the presence
of shear becomes
	\begin{equation}
	\mu_{crit}^A \approx \frac{\alpha_A A \sigma}{\pi G},
	\end{equation}
and $\alpha_A \approx 2.5$.
This surface density threshold is extremely low in the inner regions of 
galaxies with slowly rising rotation curves but approaches the
Toomre stability condition where the rotation speed is constant.


In addition to this degeneracy, many other descriptions of star
formation thresholds including gas deficiences in the outer disk
(Warmels 1988a; Cayatte \et 1990), the phase structure of the interstellar gas
(Elmegreen \& Parravao 1994), or the radiative feedback from young stars 
have been advanced (Federman \et 1979; Skillman 1987).  
With such a broad array of physical processes at work,  the K89
analysis should be repeated on a larger, more diverse sample of galaxies. 
The \Ha surface brightness profiles presented in this paper supercede those 
in the K89 paper which were based largely on HII region counts.  The \Ha images
presented alongside the threshold measurements should resolve the confusion 
which has arisen regarding what is meant by a threshold radius 
(Ferguson \et 1998). Finally, the number of galaxies with published
CO observations has greatly increased in the last decade, due in large
part to the Five College Radio Astronomy Observatory (FCRAO) 
CO survey (Young \et 1995).  Hence, it is now possible
to examine the full range of spiral galaxy types using 
atomic and molecular gas data compiled from the published literature.

The paper is organized as follows. In \S2 we describe our \Ha observations
and the atomic and molecular gas data used in our new
analysis.  In \S3 we identify the star formation thresholds, measure
gas densities at the thresholds, and evaluate the Toomre 
stability criterion.  In \S4, we discuss the failures of the model,
the shear criterion for stability, 
the influence of the stellar disk on the stability criterion,
and the implications for the global star formation rate.
The data presented here will be used to discuss 
the processes regulating the SFR in the high density regime (i.e. where
gas surface density exceeds the threshold density) in a subsequent paper. 

%
%
%
%



\section{Data}
\label{sec:data}

We selected 32 spiral galaxies divided roughly evenly between types Sab and 
Sdm that had published \co and 21-cm radial intensity profiles. Nearly half 
the galaxies are Virgo Cluster members, and 15 of the galaxies are classified 
as barred galaxies in the RC1 (de Vaucoulers system).
In the K89 paper, radial profiles were estimated for 16 galaxies from
HII region counts.  That subsample was dominated by type Sc galaxies.

Table~1 lists the type, distance, and disk inclination for each galaxy in our 
sample.  The critical density of a disk scales inversely with the assumed 
distance and is the only distance-dependent quantity that enters our analysis. 
Most of the distances come from measurements of
Cepheid variables.  We correct  surface brightness, surface density, and 
rotation speed measurements for disk inclination.  Disk inclinations and
position angle were generally adopted from 21-cm observations.  If
these values were heavily weighted by HI beyond the optical disk, we
used the shape of the optical continuum isophotes to derive a disk
inclination and position angle.  The position angles are given in Table~1
and describe the major axis of elliptical apertures used in our \Ha photometry.

\subsection{\Ha Imaging}

We use recombination line radiation from hydrogen to trace the location
of young, massive stars in the disks.  The \Ha line was
chosen for the imaging because only optical CCD cameras offered a wide enough
field of view to cover the entire disk in one or two pointings per galaxy.
The measured \Ha luminosities can be directly converted to SFR's using the calibration
of Kennicutt, Tamblyn, \& Congdon (1994) --  
	\begin{equation}
 	{\rm SFR} (\msunyr) = \frac{L_{\Ha} }
        		           {1.25 \times 10^{41}~{\rm ergs~s}^{-1}}				10^{0.4 A({\rm H}\alpha)}. 
	\end{equation} 
This relation was derived using a Salpeter initial mass function, $dN(m)/dm
 = m^{-2.35}$, for stellar masses between  m = 0.1 and 100\msun.
The stars were evolved along the evolutionary tracks of Schaller \et (1993), 
and the ionizing flux calculated from Kurucz (1992) model atmospheres.
The amount of extinction at \Ha,  $A_{H\alpha} = 0.78 A_V$ (Miller \& 
Mathews 1972), is the largest source of uncertainty in the measured SFR.

%

Narrowband continuum and \Ha images were obtained at the Steward Observatory 
Bok telescope, KPNO Burrell-Schmidt telescope, and KPNO 36-in telescope.
These CCD frames were corrected for fixed pattern noise in the standard
way using a bias frame, dome flat, and sky flat.  The 90-in data were 
obtained with a focal reducer, and scattered light produced ghost images 
which had to be removed in the background subtraction. Foreground stars 
were used to register the images and subtract the continuum emission from 
each narrowband \Ha image. The line flux was then corrected for atmospheric 
extinction and the transmission of the filter at \Ha relative to the average 
transmission over the bandpass. Data obtained at the Bok telescope and 
Burrell-Schmidt telescope were flux calibrated using observations of 
standard stars.  The five galaxies observed at the Kitt Peak 36-in and 
\n7331, \n925, \n628, and \n4402 were calibrated from the integrated fluxes 
of Kennicutt \& Kent (1983) or Kennicutt (1998).  The Burrell-Schmidt images 
do not include [NII] emission, so we increased their flux by a 
spectroscopically-determined factor ranging from 1.2 to 1.4 to consistently
include [NII] emission. Figure~\ref{fig:calibrate} compares our 
\Ha + [NII]$\lambda\lambda 6548,83$ fluxes to previously published work.
The independent measurements are typically consistent to within 15\%.

The radial distribution of \Ha emission was measured from surface photometry.
The optical nucleus of each galaxy defines the center
of a set of concentric, elliptical apertures.  The  width of the annuli
is a tradeoff between the desire to average over individual 
HII regions and the need to accurately measure breaks in the profile.  
A width of about 10\asec proved to be a good compromise for most of the 
galaxies. The profiles changed very little if this width was changed by 
a factor of two. The profiles were corrected to face-on orientation.

To convert the surface brightness profiles in \fig~\ref{fig:m1} to SFR per 
unit area,  we recommend applying an internal extinction correction, 
$A_{H\alpha} = 1.1$~mag, derived from comparison of the free-free radio 
emission and the \Ha emission from spiral galaxies (e.g. van der Hulst \et 
1988).  Correction for [NII] emission in the bandpass reduces the SFR by 15\% 
to 30\%.  Radial gradients in internal extinction are typically not strong 
compared to the scatter in extinction at a given radius (e.g. Zaritsky, 
Kennicutt, \& Huchra 1994, Webster \& Smith 1983, McCall \et 1985).  An
exception is the nuclear region.  The central star formation rates implied
by these profiles are highly uncertain due to the unknown amount of
dust obscuration,  photoionization by an active nuclei, and continuum 
subtraction errors.

The limiting surface brightness  was estimated from the uncertainty in the 
background  and the area of the aperture and then  verified by adding fake
sources to the images.  Where the covering factor of HII 
regions was low, photometry could be obtained for individual HII 
complexes using local background measurements. In other regions,
variations in the mean background level dominate the 
uncertainty in the  photometry.  In particular, scattered light in the focal 
reducer camera at the Bok telescope produced ghost images that were easily 
identified by eye but difficult to model and subtract.  We used a mean
background measurement, and then estimated the maximum error using
the full range of mean background levels on the frame. 
The errorbars shown on the radial profiles are therefore
maxima and minima in these regions.




\subsection{Molecular Gas Surface Densities}

Only a few galaxies in the sample have been completely mapped
in the \co line.  FCRAO maps have been published for \n5194 
(Lord \& Young 1990) and \n6946 (Tacconi \& Young 1989). 
 The galaxies \n5457 (Kenney \et 1991),
\n3031 (Sage \& Westphal 1991), \n5236 (Crosthsaite \& Turner in prep),
and the inner 3.2 kpc of \n2403 (Thornley \& Wilson 1995)
have been mapped with the NRAO 12m telescope.
For the rest of our sample, the CO radial profiles represent
measurements made along the major axis of each galaxy.  These profiles have 
a resolution of $\sim 45\asec$.  The data are described in detail in 
survey papers by Kenney \& Young (1988) and Young \et (1995).  

These source papers adopt CO-to-\h2 conversion factors
based on Galactic calibrations.  We follow this convention but adopt a common
scale of 
	\begin{equation}
	N(\h2) = (2.8 \times 10^{20} ~\c2)~ I_{CO} (K(T_R) \kms) 
	\end{equation} 
(Bloemen \et 1986) for all the galaxies. The beam-averaged brightness
temperature, $T_R$, inferred from the measured antenna temperture depends
on the source-beam coupling efficiency and the forward scattering
and spillover efficiency as described, for example, by Kenney \& Young (1988)
and Young \et (1995).  We adopted their calibrations, $\eta_{fss} = 0.75$
and $\eta_c = 0.73$,  for the FCRAO survey data. The \n5194 profile uses
$\eta_{fss} = 0.70$  and $\eta_c = 0.69 - 0.81$ following the 
model of Lord \& Young (1990).  The NRAO 12m source-beam coupling 
efficiency derived for M101 (Kenney \et 1991) was applied to the M81 data.  
The \h2 gas surface density was multiplied by a factor $1.4 \cos (i)$ to 
correct for inclination and to include the He mass.


\subsection{Atomic Gas Surface Densities}

The HI radial profiles were taken from 21~cm observations listed in 
Table~1. The profiles from the Warmels (1998c) paper were derived
from strip scans across each galaxy with the Westerbork Synthesis
Radio Telescope.  The half-power beam width was about 12\asec\
in right ascension by $12\asec (sin (\delta))^{-1}$ in declination.
These scans are essentially major axis profiles for many of the
galaxies, but the scans are aligned closer to the minor axis of 
\n4178, \n4321, \n4501, \n4535, \n4548, and \n4689.  

The other profiles were obtained by integrating over ellipses and 
correcting to face-on orientation.  The
typical half-power beam width of the other WRST maps is 30 - 60\asec
(Wevers 1984; Bosma 1981; Bosma 1977; Bosma \et 1981; Rots 1975).
The WSRT observations of \n5194 (Tilanus \& Allen 1991) and \n5236
(Tilanus \& Allen 1993) have higher resolution.  The surface density
profile for \n6946 comes from a Very Large Array
map (Tacconi \& Young 1986)  with 40\asec resolution. Braun \et (1994) 
describe the compilation of WRST and VLA obseravations used to construct 
the high resolution HI map of \n4826.



\subsection{Gas Kinematics}

Rotation curves are needed to calculate radial stability profiles for each 
galaxy. Our analysis uses the HI 21-cm rotation curves derived by 
Wevers (1984) and Guhathakurta (1988) and  position -- velocity diagrams 
published by Warmels (1988a, 1988b).  Velocities measured from CO observations
were included in the rotation curves for \n5194, \n5457, and \n4321 
(Kenney \& Young 1988; Young \et 1995; Sofue 1996).  These CO and HI 21-cm 
observations lack sufficient spatial resolution to determine the 
kinematics in the inner $\sim45\asec$ of the disks.
Optical emission-line velocities were used to improve inner
rotation curves of \n4548 and \n4639  (Rubin \et 1999).  In general, 
the dynamics of the inner disk are not well constrained by the
observations compiled here, and our stability analysis will
focus on the outer regions of disks.  

 
Measurements of the HI velocity dispersion are difficult to make in
galaxies that are inclined enough (relative to our sightline) to accurately
model their rotation curves.  Studies of spiral galaxies find $\sigma =
 3-10$\kms in the outer disk (van der Kruit \& Shostak 1984; Dickey \et 1990).
A similar range is estimated for dwarf irregular galaxies -- $7.6 - 11.4$\kms
(van Zee \et 1997).  The higher velocity dispersions tend to be associated with
the regions of higher star formation activity (Shostak \& van der Kruit 1984;
Jogee 1999). We adopt $\sigma = 6$\kms\ to facilitate comparison with K89.

\section{Testing the Gravitational Stability Model at the Star
Formation Thresholds}

To give the reader an appreciation of the accuracy with which
the stability criterion can be evaluated, we step through our analysis 
of \n5236.  Our \Ha atlas is then used to further illustrate what is
meant by the expression {\it  star formation threshold}.  We describe the
properties of the gas disks at the thresholds and evaluate the
local stability of the disks using the single-fluid model.


\subsection{Case Study:  \n5236}

Figure~\ref{fig:m83_gas} shows that the broken spiral pattern of \Ha emission 
from \n5236 creates a smooth radial surface brightness profile, which is well 
fitted by an exponential profile over a wide annulus. The strongest break in 
the radial profile defines the threshold radius, $\rhii = 305\asec$.  The 
azimuthally-averaged SFR there is 0.0027\msunyrk2, but the best estimate for 
the upper limit just 500~pc further out is 100 times lower. 
The \Ha surface brightness is also low in the inner disk,  where the 
star-forming regions are concentrated along the bar; but the magnitude of 
the deficit relative to the fit is much smaller.
In \fig~\ref{fig:m1}, a few HII regions are seen beyond \rhii, but their 
covering factor is much, much lower than in the region $R \le\ \rhii$.
The threshold radius derived from the azimuthally-averaged \Ha profile 
is a good approximation to the 'edge' of the star-forming disk, 
but it does overestimate the threshold radius at some position
angles (e.g. northeast side of \n5236)  since the shape of
the \Ha isophotes changes slightly with galactocentric radius.

The middle panel of \fig~\ref{fig:m83_gas} compares the radial variation in
the azimuthally-averaged HI surface density (Tilanus \& Allen 1993, TA93) 
to the variation in molecular gas surface density along the major axis 
(Young \et 1995). More recent CO mapping across \n5236 (Crosthwaite \&  
Turner, private communication) shows the molecular gas is concentrated along 
the bar near the major axis.  We find the spatial mismatch (one-dimensional 
cut vs. azimuthal averaging) exaggerates the central depletion of atomic gas 
relative to molecular gas.  The azimuthal gas distribution is more uniform 
further out in this disk.  For many galaxies in our sample, complete 
two-dimensional mapping of the gas distribution is not available.
This example draws attention to the systematic errors introduced by using
the one-dimensional cuts and suggests they are significant in the inner
disk when a gaseous bar is present.  They do not affect the measurement
of the total gas density at the threshold  radius in \n5236.  When the
\Ha surface brightness is plotted against the total gas density, a 
strong break is seen at the outer threshold.  This $\log \Sigma ~{\rm vs.}~ 
\log \mu$ relation is well described by a power law over most of the 
disk, but the spatial mismatch between the atomic and molecular gas
sampling in the inner disk artificially adds some structure to it.


The critical density for local gravitational instability, as defined
by the Toomre~Q criterion, is also shown in the middle panel of 
\fig~\ref{fig:m83_gas}.  The gas velocity dispersion is assumed to be 
constant,  so the small-scale features reflect structure in
the rotation curve of TA93 through the epicyclic frequency,
	\begin{equation}
	 \kappa^2(R) = 2(\frac{V^2}{R^2} + \frac{V}{R}\frac{dV}{dR}).
	\end{equation}
The rotation curves typically represent an average V(R) fitted to the
two-dimensional velocity field or to both sides of the major axis.
The ratio of the total gas density to the critical density,  bottom panel of 
\fig~\ref{fig:m83_gas}, falls steadily with radius beyond the bar 
region.  The spatial resolution of the rotation curve is insufficient to 
describe the critical density at radii less than $\sim 45$\asec.
 At the star-formation threshold \rhii, the gas density, 10\msunpc2, is
70\% of the critical density for instability ($\alp(\rhii)  = 0.7$),
as given in eq (2) for $\sigma$ = 6\kms.
In \n5236, the dominant error in \alp(\rhii) is the extrapolated molecular 
gas density.  If the molecular gas disk is truncated abruptly at 
$R \approx 250$\asec, then the value of \alp(\rhii) could be as low as 0.25.  
This scenario is unlikely as it would produce a
strong discontinuity in the total gas surface density.  However, 
fitting the exponential profile to only the less sensitive data 
($R_{CO} = 150\asec$ instead of $R_{CO} = 250\asec$) yields a flatter
gradient in the molecular gas density and overestimates the threshold
gas density by $\sim 40\%$.  This type of systematic error appears to
dominate the systematic error in \alp(\rhii) for a number of galaxies
in our sample.

\subsection{Threshold Radii}


\fig~\ref{fig:m1} shows \Ha + [NII] images and radial surface brightness 
profiles for the entire sample.  The threshold radii are marked.  Star 
formation thresholds, as described for \n5236, are easily identified in 27 of 
the 32 galaxies in the sample. In most of these galaxies the drop in SFR per 
unit area at the threshold is very sharp, declining by a factor of 3 to 400
over a single 10\asec resolution element.  The change in star formation 
efficiency, i.e. SFR per unit gas mass, is similar.  
The profiles for \n2903, \n4535, and \n4736 do not show such strong
discontinuities, but the sharp change in profile slope 
in the outer disks do seem to reflect some type of threshold in
the star formation law. The \Ha surface 
brightness in \n2841 and \n4698 is too low to determine whether the 
threshold effect is present.

In many disks, HII regions are seen beyond the threshold radius
(Ferguson \et 1998; Lelievre \& Roy 2000).  Their covering factor is much 
lower than it is within the main star-forming disk ($R < \rhii$).
It seems plausible that the local \mug\ / \mucrit\ ratio is  higher in these
regions.  In \n2903 for example, the HII regions beyond the threshold
radius are clearly associated with density perturbations caused by
the spiral arms.  The closer the mean gas density is to the critical
density, the more frequently we expect a given region to experience
a perturbation sufficient to bring it above the threshold.

%

\subsection{Sub-Critical Disks} \label{sec:subcrit}

Our sample contains seven sub-critical disks where
\alp never reaches the empirically-defined threshold value --
$\alpha_Q = 0.63$ in K89 and $\alpha_Q = 0.69$ in a subsequent section of 
this paper.  One member of our sample with a sub-critical disk, \n2403, 
shows widespread star formation activity, and this failure of the 
Toomre-criterion will be discussed later.  \fig~\ref{fig:Q_sub} shows the 
instability profiles for the other six galaxies with sub-critical disks.  
The gas density becomes a progressively larger fraction of the critical
density (at the radius of maximum \alp(R)) in this order --
\n4698, \n4639, \n3031, \n2841, \n4548, \n4571.

To facilitate comparison of their HII region distributions, the six
galaxies are displayed with the same intensity limits
in \fig~\ref{fig:m1}. Starting with the most stable disk in the sample, 
\n4698, only a few, small HII regions are detected.  The HII regions in 
\n3031 are confined to the spiral arms whose extent may be set by an outer 
Linblad Resonance (cf. Adler \& Westpfahl 1996, but also Westpfahl 1998).
The distribution of HII regions is much more widespread in \n2841
consistent with an instability that approaches the critical value.
In \n4548, the annulus extending from about 0.5~\rhii to \rhii is near
the instability limit, and a pair of spiral arms is present there.
The bar in \n4548 and ram pressure stripping  have strongly 
influenced the gas distribution (Vollmer \et 1999) and consequently
the star formation. This comparison indicates that --
(1) external perturbations have an important impact on where star formation 
occurs in sub-critical disks, and (2) HII regions become more prevalent as 
the instability parameter increases.  

The paucity of star formation in these six galaxies {\rm cannot}
be attributed to low  gas surface density.  The disk of \n4698 is a
special case in as much as the gas was probably removed in a collision 
with another galaxy (Valluri \et 1990). The gas surface density in 3 of the 
other 5 sub-critical disks exceeds the median threshold density (see next
section).  This result has implications beyond early-type spiral galaxies.
Low surface brightness galaxies also present a sparse distribution of
HII regions and sub-critical gas densities (van der Hulst \et 1993), 
but the paucity of HII regions can  be attributed to the low
gas surface density (only a few solar masses per square parsec).  Our
analysis strongly suggests that it is the ratio $\frac{\mug}{\mucrit}$
rather than the gas surface density alone that sets the threshold for
widespread star formation.

%

\subsection{The Relative Fraction of Atomic and Molecular Gas} \label{sec:h1frac}

Using the data described in \S~\ref{sec:data}, we measured the surface
density of atomic and molecular gas at \rhii in each galaxy.
\fig~\ref{fig:h1h2_mugas} shows the gas surface density ranges from 0.7 to 
40\msunpc2\ at the outer threshold.  In a few galaxies with prominent
gas rings, like \n4402 and \n5055, the apparent edge of the molecular
gas disk does appear to be associated with the edge of the star-forming disk.
However, in half of the galaxies in our sample,  atomic gas contributes 
most of the gas surface density at \rhii.  In eight galaxies, 
the molecular gas surface density is essentially zero at \rhii for any 
reasonable extrapolation of the CO intensity profile, and 
threshold radii predicted from the molecular gas density alone would
grossly underestimate the observed threshold radii. 
(The individual star-forming regions almost certainly contain
unresolved molecular clouds, but their contribution to the gas surface
density is apparently negligible in these regions.) Any viable threshold 
mechanism must explain the wide range in \mug(\rhii) {\it and} not depend on 
whether the gas is in atomic or molecular form.  The gravitational stability 
model meets both of these criteria.  We emphasize that it predicts 
the formation of large, self-gravitating clouds not stars.  The good 
agreement with the observed radii of the star-forming disks, however,
 strongly suggests that star formation will happen once the clouds form.


At the median threshold gas density the atomic gas fraction ranges 
from 20\% to 100\%.  Yet at densities twice as high, the ISM is almost
entirely molecular. The gas is almost entirely atomic at  surface
densities that are half the mean.   Much of the gas mass is in molecular
form in disks with high gas surface densities, including the sub-critical 
disk of \n2841.  The high atomic gas fractions in the other sub-critical
disks are consistent with their average to low gas density.
This division in atomic gas fraction is expected if the column of 
HI required to shield the molecular gas from radiation (that would dissociate 
it) is fairly uniform among disk galaxies.  The implied threshold column for
molecular gas formation ranges from 5 to 15\msunpc2. 

It is interesting to examine whether the disks that are mostly molecular
at \rhii become dominated by HI where the gas surface density drops
into the  5 to 15\msunpc2 range.  Generally, we find that no CO emission 
is detected at this radius, so the data for the outer disks are consistent 
with the same \h2/HI self-shielding column  density.
The only disk that clearly has a low
atomic gas fraction at low total gas column is \n4579. 


\subsection{The Instability Parameter at the Outer Threshold}

Critical surface densities for instability were computed from the rotation 
curves and compared to the gas densities described in the previous section.  
The median value of the instability parameter at the outer thresholds, 
$\alpha_Q = 0.69$, was found to be consistent with the K89 value.  
To reveal the physical limitations of this model, we examine the stability 
of the disks at the threshold radii in greater detail than previous work.  
Three types of second order effects are examined:  (1) Does the scatter 
in the \alp(\rhii) values reflect the accuracy of the model or the
magnitude of systematic errors?, (2) Can the sharpness of the threshold 
be predicted from the stability properties of the disk?, and (3) 
Is the stability of the gas disk influenced by the stellar disk?


\subsubsection{Disk Asymmetry}  \label{sec:assym}

The epicyclic frequency and, consequently, the critical density have
significant errors where the rotation curve is poorly determined as in
highly inclined
disks, disks with kinematic disturbances near \rhii, and disks with
offset dynamical and photometric centers.  Within our sample, the 
projection effects 
are largest for \n628 where a 1\deg\ uncertainty in the disk inclination 
implies a 20\% uncertainty in \mucrit.  The rotation curves of \n5457 and 
\n5194 are not well  defined at \rhii due to tidal disturbances, so
the threshold instability parameter is not defined in these two
galaxies.  The dynamical center of \n925 is not coincident with the 
photometric center (in \Ha, continuum, or HI) -- a situation common in 
barred Magellanic irregulars  (Pisano \et 2000 and references therein).  
In general though, uncertainty about \mucrit\ is small compared to the
error in the \mug\ measurement.

The largest systematic errors in the measurement of \mug\ are introduced by 
the asymmetry of the star-forming disks -- i.e. the lack of axial symmetry or 
roundness.  To quantify the impact on the threshold measurements, we
measured a threshold radius and stability profile along the major axis
of each galaxy.  Comparison to the azimuthally-averaged quantities
reveals the magnitude of the azimuthal variations in the instability parameter.
The largest variations are illustrated in Figure~\ref{fig:plots.pa}. We
find the following correlations between the instability
parameter and the distribution of star formation in these disks.

\begin{enumerate}


\item{
Recent star formation activity in \n4254 is skewed toward the north-eastern 
half of the disk. The threshold radius  on the western side is
only 60\% of its average value.  The instability parameter along the eastern
side of the major axis is also higher than it is on the western side. 
The threshold radius and stability parameter derived from the azimuthally 
averaged profile are heavily weighted by the eastern side of the disk.
}

\item{In \n4321, the average threshold radius \rhii provides a poor 
description of the irregular outer, \Ha isophotes.  At a position angle
of $90\deg$ for example,  the HII regions extend to a radius of 0.9\rhii 
to the west but only 0.6\rhii to the east. Inspection of \fig~\ref{fig:m1}
shows the two-arm spiral pattern also loses its definition at these radii.
The surface density is  $\sim 0.7\mucrit$ at the threshold along either 
direction of this cut because the gas density declines much more steeply 
with radius on the east side of the galaxy. In contrast, when the gas 
densities from each side of this disk are averaged together, the value of 
the instability parameter at \rhii is underestimated by a factor of 
two! Better spatial coverage drives the threshold stability parameter 
toward $\alp(\rhii) = 0.7$.
}

\item{
The star formation activity in \n4654 extends furthest to  the south-east,
but the HII regions are faint and sparse here.  The \Ha surface brightness 
is much higher on the western side of \n4654 where the star formation 
threshold is much sharper.  The gas 
surface density is sub-critical over much of the eastern side of the galaxy 
and super-critical on the western side -- consistent with the paucity
of HII regions to the east.  On the eastern side, the instability parameter
decreases more slowly with radius, however,  explaining the larger radial 
extent of the HII regions.  The sharp gradient in the instability parameter 
on the western side is likely not resolved, so the value of the instability
parameter, $\alp = 1.6$, at the western threshold is overestimated.
}

\item{In \n7331, the star formation activity is skewed toward the
southern half of the disk. The azimuthally-averaged  \Ha profile is
weighted toward this side of the disk, so the southern edge of the 
star-forming disk defines \rhii.  Along the southern side of the major
axis, the gas density is 0.8 \mucrit at \rhii.  At this radius, 
$\alp = 0.4$ on the northern side of the disk. However, at the smaller radius
defined by the northern edge to the star-forming disk, the gas density is 
0.6\mucrit. In this case the azimuthally-averaged value of the
instability parameter, $\alp(\rhii)=0.4$, clearly underestimates the true
value at the edge of the star-forming disk. 
}
\end{enumerate}

We have demonstrated that using azimuthally-averaged gas densities and SFR's
can lead to errors in \alp(\rhii) as large as a factor of two when
the disk is highly non-axisymmetric. The theoretical basis for the Toomre 
stability criterion breaks down in these irregular systems, so it is 
noteworthy that a local Q criterion appears to be so useful in describing 
the distribution of star formation in these highly disturbed disks.  The
examples presented demonstrate that the distribution of star-forming regions 
clearly becomes much more sparse, similar to the subcritical disks, in 
regions where the instability parameter is low.   The threshold
values of the instability parameter also show a tendency to converge
when better spatial coverage is used.

\subsubsection{Star Formation Beyond the Threshold Radius} 

Gravitational threshold models only predict a sharp edge to the
star-forming disk if the gas disk is perfectly smooth.  The structure 
of the interstellar medium in real galaxies must modify
this picture (Kennicutt 1989).  Non-axisymmetric modes and local 
perturbations generate structure on large and small scales, respectively.
Such regions may allow cloud formation (and subsequently star formation)
even where the azimuthally-averaged gas density is sub-critical.
For example,
in \S~\ref{sec:subcrit}, the spiral density waves in the subcritical
disks of \n3031 and \n4548 are supported where the gas surface density
exceeds 0.3\mucrit rather than 0.69\mucrit.  The lower threshold is consistent
with models of swing amplification.  The more isolated HII regions in
the subcritical disks are generated by a different mechanism.  We suggest
that random density perturbations generate some super-critical regions
in disks that have surface densities close to the critical density for
instability.  Support for this viewpoint comes from the sub-critical 
disks. Isolated HII regions were found most frequently when the gas surface
density was only slightly sub-critical,  and no HII regions were found
where the gas surface density was less than 10\% of the critical 
density.  As the average gas surface density approaches the critical density,
the amount of star formation activity in isolated, star-forming regions 
clearly increases.
Could a combination of local perturbations and spiral arms also explain
the existence of HII regions at radii beyond the threshold radius?
If so, star formation activity should be more prevalent where the 
\mug/\mucrit ratio remains high (i.e. greater than 0.3) beyond \rhii.

To test this hypothesis, we need to minimize
the systematic errors caused by disk asymmetry at \rhii, so we compare
the 11 galaxies with the smallest azimuthal variations in threshold radius.
\fig~\ref{fig:Q_edge} shows the radial stability profiles for ten
members of our sample; the profile for \n5236 was shown previously.
The spread in the instability parameter is only a factor of two 
at the threshold radius where the profiles converge. Beyond the
threshold radius, the shallowest gradient is found in the disk of \n2403 
which remains marginally stable. Not surprisingly, many HII regions are seen 
beyond \rhii in \n2403.  The gas surface density in \n6946 also remains close
to the critical density, i.e. $\mug\ \ge\ 0.4\mucrit$, out to 2\rhii; and a
large number of HII regions are seen beyond \rhii (cf. Ferguson \et 1998).
In contrast, the steepest gradients in the instability parameter,  
	\begin{equation}
	\mid \frac{d(\mug/\mucrit)}{dR}\mid\ > ~0.1~{\rm kpc}^{-1}, 
	\end{equation}
are found in  \n4402, \n4713, and \n5236. 
None of these galaxies show much \Ha emission beyond the threshold radius.  
The instability parameter is also low outside the threshold radius in 
\n2903, but the star formation  is confined to the spiral arms.
locally.  The likelihood of finding isolated HII regions outside \rhii 
therefore appears to scale with the value of the instability parameter.

It is interesting to examine why, beyond \rhii, $\alp(R)$
 remains high in some disks
and drops sharply in others. \fig~\ref{fig:sd_edge} shows the gas surface 
density profiles for the same ten galaxies.  The HI disks of
the Virgo cluster galaxies \n4394, \n4402, and \n4689 appear to have been
stripped at radii $R \ge \rhii$.  When the molecular gas density is included
the difference in total gas surface density relative to the field galaxies is 
reduced, but the Virgo members still tend to have lower values of
of the instability parameter outside \rhii.  Inspection of \fig~\ref{fig:m1}
shows that the Virgo galaxies show HII regions beyond \rhii less
frequently than the field sample.

\subsubsection{Variations in the Threshold Stability Parameter with
Galaxy Type}

\fig~\ref{fig:alpha} shows our best estimates of \alp(\rhii)
at the outer edges of 26 disks.\footnote{
	The six galaxies with sub-critical disks are omitted.
	\n5194 and \n5457 are not shown since the rotation curves are
	not well-defined at \rhii.}
The significance of the weak trend with galaxy type is difficult to assess
quantitatively.  The error bars represent limiting values and are set by the
magnitude of systematic errors associated with one-dimensional
gas profiles in non-axisymmetric disks, tidal disturbances,
the extrapolation of the CO intensity profile to larger radii, and disk 
inclination.  The dominant error terms differ in the early-type and late-type 
disks, so we discuss their \alp(\rhii) values separately.

The earlier type disks (Sa to Sb) in \fig~\ref{fig:alpha} 
tend to have \alp(\rhii)
values lower than the sample median.  The discrepancy is not significant
for \n2903, \n4321, and \n7731.  The range of \alp values shown for each
of them is consistent with the sample median, 0.69, because corrections for 
non-axisymmetric structures were shown to drive \alp in this direction 
(S~\ref{sec:assym}).   The \alp(\rhii) values for the early-type 
galaxies \n4394, \n4402, \n4579, and \n4689 range from 0.5 to 0.6 and
depend on extrapolations of the CO intensity profile.  
In all of the galaxies where molecular gas dominates the gas density
at the threshold, \alp is less than or equal to 0.69; and most of  these
galaxies are type Sab-Sbc.  It is unlikely that the gas surface density
is systematically underestimated in these disks because the
two largest biases in interpretation of the CO measurements work to
increase the molecular gas density.  First, since the CO emission 
is not well-resolved, particularly in \n4394, \n4402, and \n4689, 
the tendency is to overestimate scale lengths and artificially raise the
molecular gas surface density at the threshold radii.
Second, the CO/H$_2$ conversion factor is expected to be lower in disks with
low metallicity.  If the early-type spirals are more evolved
than the late-type spirals, i.e. have consumed a larger fraction of
their original gas content, then they should have higher metallicity on
average.  Hence, we have no reason to expect a low CO/H$_2$ conversion 
factor in the early-type galaxies.  It is plausible then that the
$\mug/\mucrit(\rhii)$ ratios  are somewhat lower than the mean 
in these early-type disks.

In contrast, the threshold values of the instability parameter
 for the HI-dominated edges scatter to both sides 
of 0.69. Lower mean metallicities in the later-type disks would introduce an 
artificial negative slope in \fig~\ref{fig:alpha}, which is not seen.  We 
find no trend toward higher values of the instability parameter in late-type
galaxies.  In \n628, the estimated threshold value of the 
instability parameter, $\alp(\rhii) = 0.86$, is consistent with the 
mean, $\alp _Q = 0.69$, given the uncertainty in the rotation speed described
in \S~\ref{sec:assym}.  In \n925, the 
instability parameter is overestimated on the western side of the
galaxy when the offset between the photometric and dynamical centers
of this galaxy are neglected.  (The kinematic radius is smaller than the \Ha 
radius, so $\kappa$ is higher than we estimated assuming common 
photometric/dynamic centers.)    A detailed analysis using  
$\sigma _g = 10$\kms found \alp is 0.3 at the threshold (Pisano \et 2000) 
which corresponds to  $\alp = 0.5$ for our assumed value of $\sigma = 6$\kms.  
Evaluation of the instability is particularly difficult in these late-type
spirals, owing to the non-axisymmetric structures. We find no trend toward 
lower \alp(\rhii) in the low angular velocity galaxies in our sample, all of 
type  Scd to the Sdm, as one might have expected based on the low  \alp 
values reported in dwarf irregular galaxies (HEB98). 

%

\subsubsection{Conclusions}
The asymmetries of many disks in the sample compromise the accuracy
of the stability analysis.  We have demonstrated, however, that
when disks are examined along position angle cuts that eliminate this 
uncertainty, the convergence of the stability profiles at the threshold 
radii is tighter.  Outside the threshold radius, the disks with gas 
densities well below the critical density do not show many HII regions 
while those with prominent populations of HII regions tend to be
near the threshold surface density.  The weak trend toward lower
values of $\alpha_Q$ in earlier type spiral galaxies is consistent
with these disks having a larger stellar mass fraction (e.g. 
Jog \& Solomon 1984).

%
%
%

\subsection{Inner Disk Thresholds}

We also examined whether simple gravitational stability arguments
might be as successful describing star formation thresholds in the inner 
regions of disks as we have shown them to be at the outer edge of 
star forming disks.

The sharpest star formation thresholds in the inner disk are associated
with rings of intense star formation activity -- e.g. see \n4569, \n4579, 
\n4639, and \n4736 in \fig~\ref{fig:m1}.  Does the instability parameter,
$\alp(r)$, increase across these rings?  The beamwidth of the HI and CO 
observations is inadequate to evaluate the instability parameter across the 
rings in \n4569 and \n4639.  The major axis instability profile across
\n4579 reaches $\alp = 0.6$ on the western side of the ring, but 
the gas density along the eastern major axis is much lower than the critical 
density.   The ring of HII regions in \n4579 is brighter on the west 
side than the east side, so  star formation in the ring appears to be 
correlated with higher values of the instability parameter, i.e. 
lower Q values.  In \n4736, the ring of HII regions at $R \sim 45$\asec 
coincides with an inner Lindblad resonance (Gerin \et 1991; Gu \et 1996; 
Wong \& Blitz 2000).  The spatial resolution of the gas data does not resolve 
the 680~pc wide ring, but the lower limits on the gas surface density
in the ring are sufficient to raise the instability parameter to
values of 0.6 to 1.0. near the inner edge of the ring ($R \approx 34\asec$).
Higher resolution data indicate $Q < 1$ in localized regions of the ring
and $Q \sim 3$ averaged across the ring (Wong \& Blitz 2000).


A few galaxies without rings have gas surface densities less than 
0.69\mucrit in their inner disks. We show the \alp(R) profiles for
the three with the best spatial resolution in Figure~\ref{fig:Q_in}.  
We expect to see little \Ha emission at radii less than the inner threshold
radii in these galaxies. In \fig~\ref{fig:m1}, 
the disks of \n628 and \n4535 show little
star formation inside the predicted threshold radii of 70\asec 
and 50\asec, respectively.  Star formation in the central region of 
\n4178 is confined to the bar, where the gas density is likely higher 
than the azimuthal average, so the star-forming regions may not be
sub-critical.  The inner thresholds in these three galaxies are
therefore consistent with the gravitational threshold model.  Only
the star formation in the central region of \n2403 is in direct conflict 
with the gravitational threshold model.

The inner regions of many spiral galaxies have a deficit of \Ha emission, 
relative to a fitted exponential disk  -- e.g.
\n628, \n925, \n2403,  \n3031, \n4321, \n4394,  \n4535,  and \n5236.  
Aside from the cases already discussed and \n4394, these breaks in the 
\Ha surface brightness profile are not large enough to be called star 
formation thresholds.  This property of the \Ha surface brightness profiles 
was noticed by Hodge \& Kennicutt (1983), but the physical cause has not 
been established.  The region of low SFR is clearly associated with a bar in 
\n5236, \n4394, and \n925.  The inner threshold in \n3031 coincides with the 
inner Linblad resonance. Many galaxies with central \Ha deficits 
have sub-critical gas surface densities in the inner disk, and those
with supercritical gas densities have bars or resonances. The only galaxy
in this list with a supercritical gas density in the center and no identified
bar or resonance is \n4321.  Apparently, bars and resonanaces can inhibit 
star formation in the central regions of disks that are unstable to 
axisymmetric perturbations.

\section{Discussion} \label{sec:discuss}

Our results support previous studies which have claimed that
the outer edges of spiral disks are well described by the
gravitational stability of a single fluid, isothermal gas
disk (Kennicutt 1989).  We have shown that this criterion is often 
a good description of star formation thresholds in the inner disk as well.
The aim of this section is to gain further insight into the 
underlying physical processes that regulate disk star formation.
We examine the galaxies where the gravitational 
instabilty model fails and evaluate the utility of a qualitatively 
different threshold condition. A clear prediction of the gravitational
threshold scenario is that the stellar disks should affect $\alpha_Q$.
We examine whether first-order accounting of the stellar disk changes
the scatter in the distribution of \alp(\rhii) measurements.  Finally,
the impact of these star formation formation thresholds on the global
star formation rate is summarized.

\subsection{Shear Models}

\subsubsection{\n2403 and M~33}

The gas density in the inner 240\asec of \n2403 is below the
critical value and less than half the critical density within
150\asec\ of the center of the disk. Yet the inner half of the
disk is covered with HII regions.  No other
galaxy in our sample presents widespread star formation in regions 
where the gas disk is predicted to be stable.  This
contradiction, and the very similar situation in M~33, were noticed
by Kennicutt (1989).  The most obvious explanations have been dismissed  
by Thornley \& Wilson (1995).  In particular, if the H$_2$/CO conversion
factor were as much as five times higher than the Milky Way value, the
implied increase in the molecular gas density would bring the total gas
density up to the critical density in the inner disk  without affecting
the good agreement at the outer threshold.  Comparison of virial and
molecular masses for clouds in M~33, however, indicate the conversion
factor is similar to the Galactic one (Wilson 1995).  The metallicity
and rotation speed of M~33 are similar to \n2403, so Thornley \& Wilson
find no reason to suspect a higher surface density of molecular gas in
either galaxy.

The star-forming disks in many dwarf irregular galaxies also appear to be 
sub-critical when the Toomre~Q parameter is used to estimate the critical 
density, and it has been suggested that the shear criterion, Eqn. 5, 
describes the critical density better than the Toomre~Q criterion 
(Hunter \& Plummer 1996; HEB98).  The disks of \n2403 and M~33 share an 
important dynamical property with dwarf irregular galaxies.  The shear rate 
is low in the inner disk where the rotation speed is rising slowly with 
radius.  The disk stability analysis for the irregulars, while quite
interesting, is not yet compelling.  Large systematic uncertainties
arise from the highly irregular shape of the disks, greater uncertainty 
about the CO/H$_2$ conversion factor at very low metallicities, and
the assumption that the molecular/atomic gas fraction is constant with
radius.  

Figure~\ref{fig:shear_n2403} compares the shear criterion for the critical
density to the Toomre critical density across \n2403 and M~33.  
The gas surface densities for M33 data come from Newton (1980) and 
Young \et (1995), and the threshold radius for star formation is 29\farcm\  
In both spiral galaxies, the gas surface density exceeds $\mu_{crit}^{A}$ 
but falls below the Toomre critical density.  The presence of widespread
star formation can be interpreted as evidence that cloud formation is 
limited by the time available for cloud growth (in the presence of shear)
rather than gravitational instability.  It remains unclear whether this
explanation is correct.  One disk in our sample, e.g. \n4535, has an inner 
threshold which is well-described by the Q~criterion but not the shear 
criterion.  Also, models of self-regulated star formation predict $Q \sim 1$ 
throughout the disk (Silk 1997).  Departure from this equilibrium condition 
may indicate the disk is in a transient state.  The unusually high infall 
rate of neutral hydrogen onto the \n2403 disk, and perhaps M~33 as well, 
support this view (Schaap, Sancisi, \& Swaters 2000; Sancisi \et 2000).

\subsubsection{Shear Criterion at the Outer Threshold}

The galaxies in our sample generally have flat rotation curves at the 
threshold radius, so the  critical density based on the Toomre~Q criterion
reduces to 
	\begin{equation}
	 \mucrit = \alpha_Q \frac{1.41 \sigma \Omega}{\pi G}. 
	\end{equation}
The shear criterion for the critical density has the same radial dependence
where the rotation speed is constant, 
	\begin{equation}
	 \mucrit _{A} =  \alpha_A \frac{0.5 \sigma \Omega}{\pi G}. 
	\end{equation}
Since the fitted value of $\alpha_Q = 0.69$, we need $\alpha_A =  2.0$
to obtain $\mug = \mucrit _{A}$ at \rhii\ for the disks with constant
rotation speed at \rhii.  This normalization factor is very close to
that predicted by Elmegreen (Elmegreen 1993; Hunter, Elmegreen, \& Baker
1998).  If the rotation speed is constant at the outer threshold, then no 
distinction can be made between these two criteria for the outer edge of a 
disk.

The galaxies in our sample with the largest gradients in rotation speed 
at \rhii are \n4254, \n4394, \n4402, \n4535, and \n4569. The gradient
in \n4394 is highly uncertain due to the low resolution of the HI position -- 
velocity diagram for \n4394.  \fig~\ref{fig:Q_shear} compares the Toomre
(top panel) and shear (bottom panel) representations of the critical density 
for the other four galaxies.  The scatter among the values of \mug/\mucrit\
is actually smaller for the shear model than the Toomre~Q model.  However,
when the $\alpha _A$ term in the shear criterion is normalized using 
the galaxies with flat rotation curves at \rhii, as represented by 
\n2403 and M~33 in \fig~\ref{fig:Q_shear}, the scatter in the
\mug/\mucrit(\rhii) values exceeds that obtained with the Toomre~Q criterion.
The shear criterion does not appear to be as robust an indicator of the 
edges of star-forming disks in spiral galaxies as the Toomre criterion.  
The shear model was also unable to adequately describe the 
edges of the disks in dwarf irregular galaxies (Hunter \et 1998).


\subsection{The Influence of Stellar Disks on Instability}

Real galaxies contain both gas and stars, and the system can be
unstable even when both the stellar and gaseous components
individually meet the requirements for stability (Jog \& Solomon 1984;
Elmegreen 1995; Wang \& Silk 1994).  Wang \& Silk give an effective Q
parameter for this two-fluid instability such that $Q_{\rm eff} =
\alpha_{eff} Q$, where 
\begin{equation} 
\alpha_{eff} = (1 + \mu_* \sigma_g /
(\mu_g \sigma_*))^{-1}.  
\end{equation} 
In the limit of a low stellar
surface density or large stellar velocity dispersion, we have $\alpha_{eff}
\approx 1.0$; and $\alpha_{eff}$ gets smaller as the stellar disk becomes
more unstable relative to the gaseous disk. Following Wang \& Silk,
the parameters for the Milky Way disk yield $\alpha_{eff} = 0.72$ in the
solar neighborhood.

How much should we expect this parameter to vary over our sample?
The gas disk is a smaller fraction of the total gas mass in the
earlier type galaxies, so the stellar disks are expected to have a more
important influence on the stability of those disks.  
Scaling to the disk of the Milky Way (e.g. Kenney \& Young 1989),  
the measured rotation velocities of these disks at \rhii imply average
surface densities of 15 to 300 \msunpc2.  If the other parameters
are held at their solar neighborhood values, then the highest surface
density disks should become unstable at a value of \mug/\mucrit that
is $\sim 25\%$ lower than $\alpha_Q$ in the lower density disks.  In other
words, we would not be surprised to see $\alpha_Q$ values ranging from
0.25 to 1.0 across our sample.  Better measurements of the stellar
velocity dispersion and surface density would be needed to test
whether the lack of such a trend posses a problem for the gravitational
instability model.


\subsection{Consequences for Star Formation Prescriptions}

We have presented a picture where a single parameter, the 
gas surface density relative to the critical density for gravitational
instability, determines whether  the disk forms stars.  Evolution in 
the cosmic star formation rate is likely driven, at least in part,
by the evolution in the rate of galaxy -- galaxy interactions with lookback
time (e.g. Tan \et 1999). In our star formation recipe, the environment 
influences the SFR indirectly through its impact on the large-scale gas 
distribution. Including the star formation threshold in the recipe allows 
the SFR to react non-linearly to small changes in the gas distribution.

Our \Ha atlas demonstrates  that the distribution of star-forming regions 
in many disks is better described as {\it asymmetric} rather than
{\it axisymmetric}. The correlations we
find between local values of the instability parameter and the distribution
of HII regions clearly indicate that the Toomre~Q parameter provides useful
guidance about the location of star forming regions in these disks.  For
example, in \S~\ref{sec:assym}, we found that disks are more unstable
on the side with the higher star formation rate.  Prior to that
we showed that star formation at sub-critical radii in
\n2903, \n3031, and \n4548 is confined to the spiral arms. The average gas 
surface density was within a factor of three of the critical value in all
three cases.  The increase in gas surface density in the arms is expected to 
be larger than the increase in the 
epicyclic frequencey $\kappa$ (and hence critical
density) caused by unresolved, streaming motions (Rand 1993; Elmegreen 1994), 
so the arm gas may be supercritical locally.  Since these features will not
be resolved in cosmological simulations either, it would be reasonable to
lower the $\alpha _Q$ value in Eqn.~2 from 0.69 to $\sim 0.3$ in galaxies
undergoing weak interactions to account for unresolved, super-critical
structure in the gas distribution

The fraction of gas residing in sub-critical regions of disks varies
widely among the galaxies in our sample.  Accounting for gas beyond the 
threshold radius produces the most significant correction to the global 
star formation rate.  Among the spiral galaxies in our sample, the gas mass 
interior to the threshold radius \rhii varies from 0.14 to 0.94 of the total 
gas mass.  Omitting the threshold radius from a model therefore leaves
the global star formation rate uncertain by a factor of $\sim 7$.  The median
value of the gas fraction interior to \rhii is 0.6.  
Ram pressure stripping has removed diffuse, 
atomic gas from at least these Virgo cluster galaxies
-- e.g. \n4394, \n4402, \n4569, and \n4689
(Warmels 1988a, 1988b; Giovanelli \& Haynes 1983; Kenney \& Young 1989);
yet a histogram of the gas mass fraction within \rhii for the Virgo subsample
is indistinguishable from the histogram for the field subsample.

\section{Summary}

Using a larger sample of spiral galaxies than previous studies, we
tested the thesis that widespread star formation occurs where the 
gas disk is unstable to axisymmetric perturbations.  We located 
the edge of the star-forming disk using our new \Ha photometry, measured the 
gas surface density there, and derived the critical surface density for 
gravitational instability using the Toomre~Q criterion. Our results confirm
previous work by Kennicutt (1989) which found the threshold gas density 
varied by at least an order of magnitude among spiral galaxies but that 
the ratio of the gas density to the critical density was much more uniform. 
Among 26 galaxies with well-defined thresholds, we found the median ratio of 
gas surface density to critical density is $\alp _Q = 0.69 \pm 0.2$, 
with \alp(\rhii) defined by eq.~(2) and an assumed gas velocity dispersion of 
6\kms.  At the thresholds, the gas is primarily atomic (molecular) in the disks
with the lowest (highest) gas surface density.  These results confirm the
surprising accuracy with which a simple stability criterion describes the
extent of the star-forming disk in large, objectively-defined samples of 
galaxies.

This work also exposes the limitations of that simple model.  The
azimuthally-averaged, threshold values of the instability parameter, 
\alp(\rhii),
range from 0.3 to 1.2.  A large part of that dispersion is due to
uncertainties associated with asymmetric gas distributions, asymmetric mass 
distributions, and the extrapolation of the molecular gas surface density 
profiles.  We demonstrated that the scatter in \alp(\rhii) values is
significantly reduced when \rhii and \alp(\rhii) are fitted as functions
of position angle.  This result demonstrates that the simple Toomre~Q
stability criterion is useful for describing the distribution of star forming 
regions even in disks with quite non-axisymmetric gas distributions.
The  weak trend toward lower values of \alp(\rhii) in Sab-Sbc
galaxies, however, is not easily explained by any of the above systematics.
These early-type spiral galaxies tend to have more massive stellar disks, so 
the lower values of $\alpha_Q$ would be consistent with the predicted 
de-stabilizing effect of a stellar disk on the gas disk.  
We find several sharp thresholds in the inner disks that are well described
by  gravitational stability alone. Star formation is also supressed
in the centers of a few disks where the gas density is super-critical, but
each of these disks has a strong stellar bar.  More detailed analysis
of the star formation rate in the inner region of disks requires higher 
spatial resolution (cf. Jogee 1999; Jogee \et 2001).  We conclude that
streaming motions in the disk and the gravitational attracton of the 
stellar disk require small corrections to the simple instability
parameter. Much of the uncertainty in current estimates of the instability
parameter, however, should be resolved with complete two-dimensional 
mapping of the gas surface density and velocity field (e.g. Thornley
\et 1999).  

One clear failure
of the gravitational threshold model remains two sub-critical
disks with widespread star formation in the inner disk -- \n2403 and M~33.  
Both are low mass disks with low shear rates 
(i.e. slowly rising rotation curves).
Elmegreen (1993) and HEB98 have suggested that star formation is 
limited by the time available for clouds to grow via inelastic collisions
(in the destructive presence of disk shear) rather than cloud formation, 
which they argue happens easily via magnetic instabilities.  We demonstrate 
that their shear criterion offers an explanation for the high star 
formation rates in the inner disks of \n2403 and M~33 but find it does not 
consistently predict the outer (or inner) threshold radii.
The vertical structure of these disks is also unusual in that
the infall rate is high, and the high star formation rates may
be indirectly related to this transitional state.
Since the star formation rate depends more naturally on  gas volume density 
rather than surface density, the impact of the vertical structure of gas 
disks on the star formation law remains an important issue in a 
more general sense.  Unfortunately, radial variation in the gas velocity 
dispersion remains controversial due to the difficulty of the measurement
(e.g. Ferguson \et 1998; Sellwood \& Balbus 1999).  Among the galaxies
in our sample, the intrinsic dispersion in the threshold value of the 
instability parameter is small for a constant velocity dispersion leaving
little motivation for strong variation in velocity dispersion with radius.
Given the difficulties in measuring velocity dispersions accurately, the
issue could be settled by comparing the threshold radii reported here
to models of the transition radius from disk self-gravity to halo 
self-gravity.

The utility of the Toomre~Q criterion extends beyond axisymmetric 
perturbations.
Previous studies of sub-critical disks have focused on low surface
brightness galaxies, which also have lower than average gas surface density.
We showed that star-formation is similarly suppressed in disks with normal 
gas surface density if the rotation speed is fast enough to stabilize
the disk.  As the gas density approaches the critical density for instability,
more isolated HII regions appear in the sub-critical disks.  Similarly, in
star-forming disks, HII regions are found outside the threshold radius 
most frequently when the gas density in the outer disk is near (i.e. within 
a factor of three roughly) the critical density.  Star formation is
sometimes confined to a spiral density wave in these sub-critical
environments, and the co-rotation radius of the spiral pattern may describe 
the outer edge of the star-formation activity.
This result seems to support
the view advanced by Rudnick, Rix, \& Kennicutt (2000) that tidal
interactions impart a significant boost to the star formation rate in
disk galaxies.

Star formation thresholds in the outer disk change one's view of disk
star formation in two ways.  First, in a typical disk today, roughly 40\%
of the gas mass is not included in the empirical relation that defines
the global Schmidt law for the SFR, and the analogous
correction should be made when global star formation laws are applied
in simulations.  Second, disks are clearly 
not perfectly self-regulated if their outer disks are not forming stars.
Significant velocity dispersions in these regions imply a physical process 
other than feedback from massive stars contributes to disk
heating  (Sellwood \& Balbus 1999).  
Because of the threshold effect in the star formation law, 
the central density and angular momentum of the halo are clearly of
fundamental importance to the star formation history of a disk galaxy
(Dalcanton, Spergel, \& Summers 1997; Mo \et 1998; Mao \et 1998)

%

\acknowledgements
{Acknowledgements: We thank Rene Walterbos for making the KPNO 36-in data
available and Jean Turner \& Pat Crosthwaite for providing a CO map of 
\n5236 in advance of publication.  We thank Bruce Elmegreen, Sharda Jogee,
and Nick Scoville for their comments on a draft of the paper and enlightening
discussions.  The comments of an anonymous referee were also appreciated.
  This research has made use of the NASA/IPAC
Extragalactic Database (NED) which is operated by the Jet Propulsion 
Laboratory, California Institute of Technology, under contract with the 
National Aeronautics and Space Administration. 
CLM acknowledges support from a Sherman Fairchild fellowship.
RCK gratefully acknowledges the support of NSF Grant AST-9900789.
}


%
\footnotesize
\begin{table}
\caption{Data}
\begin{tabular}{lllllll}
\hline
\hline
Galaxy	& Type\tablenotemark{d}
		& d(Mpc) 
		& $\iota$\tablenotemark{e} 
		& PA(\deg)
		& \Ha\tablenotemark{a} 
		& Reference (d, RC, HI, CO) \\
\hline
\n628	& SA(s)c
		& 11.4\tablenotemark{b}
		& 5\deg
		& 25
		& K36
		& 1, 2 and 3, 1 \\
\n925	& SAB(rs)cd
		& 9.3
		& 55.2\deg
		& 102
		& S90
		& 4, 2 and 5, 2, 1 \\
\n2403  & SAB(s)cd   
		& 3.6
		& 60\deg
		& 125
		& BS     
		& 6, 2 and 5, 2, 1 \\
\n2841  & SA(r)b    
		& 10.0\tablenotemark{b}
		& 68\deg
		& 148
		& K36
		& 1, 7, 7, 1 \\
\n2903	& SAB(rs)bc	
		& 6.6\tablenotemark{b} 
		& 60.0\deg
		& 44
		& S90
		& 1, 2, 2, 1 \\
\n3031	& SA(s)ab	
		& 3.6
		& 58.7\deg
		& 149
		& BS
		& 8, 9, 10, 11 \\
\n4178 	& Sdm	
		& 17.0
		& 69.0\deg
		& 30
		& S90 
		& 12, 13, 14, 15 \\
\n4254	& SA(s)c	
		& 17.0 
		& 28\deg
		& 56
		& S90
		& 12, 13, 14, 15 \\
\n4321	& SAB(s)bc	
		& 16.1
		& 30\deg 
		& 120
		& S90
		& 16, 17, 14, 15 \\		
\n4394	& (R)SB(r)b	
		& 17.0
		& 25\deg
		& 108
		& S90
		& 12, 18, 14, 15 \\
\n4402	& Sb	
		& 17.0
		& 75.0\deg
		& 90
		& S90 
		& 12, 13, 14, 15 \\
\n4501	& SA(rs)b	
		& 17.0
		& 58.0\deg
		& 140
		& K36
		& 12, 13, 14, 15 \\
\n4535	& SAB(s)c	
		& 16.3
		& 44\deg
		& 0
		& S90
		& 19, 13, 14, 15 \\
\n4548	& SBb(sr)	
		& 16.1
		& 33\deg
		& 136
		& S90
		& 6, 37, 14, 15 \\
\n4569	& SAB(rs)ab	
		& 17.0
		& 65\deg
		& 19
		& S90
		& 12, 13, 14, 15 \\
\n4571	& SA(r)cd	
		& 17.0
		& 28\deg
		& 55
		& S90
		& 12, 20, 14, 15 \\
\n4579	& SAB(rs)b	
		& 17.0 	
		& 36\deg
		& 95
		& S90
		& 12, 13, 14, 15 \\
\n4639	& Sbc	
		& 17.0 
		& 45\deg
		& 123
		& S90
		& 12, 37, 14, 15 \\
\n4647	& SAB(rs)c	
		& 17.0
		& 37\deg
		& 110
		& S90
		& 12, 20, 14, 15 \\
\n4651	& SA(rs)c	
		& 17.0
		& 42\deg
		& 71
		& S90
		& 12, 18, 14, 15 \\
\n4654	& SAB(rs)cd	
		& 17.0
		& 52\deg
		& 128
		& S90
		& 12, 13, 14, 15 \\
\n4689	& SA(rs)bc   
		& 17.0
		& 30\deg
		& 165
		& S90
		& 12, 13, 14, 15 \\
\n4698	& SA(s)ab	
		& 17.0
		& 57\deg
		& 164
		& S90
		& 12, 18, 14, 15 \\
\n4713	& SAB(rs)d	
		& 17.0
		& 49.0\deg
		& 100
		& K36
		& 12, 20, 14, 15 \\
\n4736  & (R)SA(r)ab   
		& 4.7\tablenotemark{b}
		& 35\deg
		& 122
		& K36
		& 1, 21, 21, 1 \\
\n4826  & (R)SA(rs)ab	
		& 5.4\tablenotemark{b}
		& 52\deg
		& 112
		& S90
		& 1, 22, 22, 1 \\
\n5055	& SA(rs)bc	
		& 8.4
		& 55.2\deg
		& 99
		& K36
		& 1, 2, 2, 1 \\
\n5194  & SA(s)bc pec  
		& 8.1
		& 20\deg
		& 170
		& BS
		& 1, 23, 24, 25 \\
\n5236	& SAB(s)c	
		& 4.1\tablenotemark{c}
		& 23.9\deg
		& 225
		& BS
		& 26, 27, 27, 1 (36)\\
\n5457  & SAB(rs)cd   
		& 7.4
		& 18\deg
		& 44
		& BS
		& 28, 29 and 30, 30, 29 \\
\n6946	& SAB(rs)cd	
		& 4.8
		& 30.0\deg
		& 62
		& BS
		& 21, 32, 33, 34 \\
\n7331	& SA(s)bc	
		& 15.1
		& 74.9\deg
		& 167
		& S90
		& 35, 7, 7, 1 \\ 
%
\hline
\end{tabular}
%
%
\tablenotetext{a}{Note. -- Telescope used for \Ha observation:
		  Burrell-Schmidt (BS); 
		  KPNO 36-in (K36);
		  Steward 90-in (S90).
		  }
\tablenotetext{b}{Published distance adjusted to $H_0 = 70$km s$^{-1}$ Mpc$^{-1}$.}
\tablenotetext{c}{Distance to M83 group.}  
\tablenotetext{d}{deVaucouleurs type from RC1}
\tablenotetext{e}{Inclination of the disk such that a face-on orientation
is 0 degrees.}
\tablerefs{
	(1) Young \et 1995;
	(2) Wevers \et 1986; 
	(3) Kamphuis \& Briggs 1992;
	(4) Silbermann \et 1996; 
	(5) Pisano, Wilcots, \& Elmegreen 1998; 
	(6) Graham \et 1998;
	(7) Bosma 1981;   
	(8) Freedman \et 1994b;
	(9) Adler \& Westpfahl 1996;
	(10) Rots 1975;
	(11) Sage \& Westpfahl 1991;
	(12) Freedman \et 1994a;  
	(13) Guhathakurta \et 1988;
	(14) Warmels 1988c; 
	(15) Kenney \& Young 1988;
	(16) Ferrarese, L. \et 1996, \apj, 464, 568.
	(17) Sofue 1997; 
	(18) Warmels 1988a;
	(19) Macri \et 1998;
	(20) Warmels 1988b; 
	(21) Bosma 1977; 
	(22) Braun \et 1994;
	(23) Sofue 1996; 
	(24) Tilanus \& Allen 1991; 
	(25) Lord \& Young 1990; 
	(26) Saha \et 1995;
	(27) Tilanus \& Allen 1993; 
	(28) Kelson \et 1996; 
	(29) Kenney \et 1991; 
	(30) Bosma, Goss, \& Allen 1981; 
	(31) Sharina \et 1997; 
	(32) Carignan \et 1990; 
	(33) Tacconi \& Young 1986; 
	(34) Tacconi \& Young 1989; 
	(35) Hughes \et 1998;
	(36) Crosthwaite \et 2000 (in prep);
	(37) Rubin \et 1999.
	}
\normalsize
\end{table}



\clearpage
FIGURE CAPTIONS

	\begin{figure}[h]
	\caption{
	Integrated \Ha + [NII] fluxes vs published values of
	Kennicutt \& Kent (1983) and Kennicutt (1998). Open symbols denote 
	galaxies with dependent calibrations -- i.e. the data presented
	here were calibrated using the data plotted along the abscissa.
	}
	\label{fig:calibrate} \end{figure}
	\begin{figure}[h]
	\caption{
	(top) The \Ha surface brightness profile for \n5236. 
	The dashed line is the fitted exponential disk.  The threshold 
	radius is shown by the 
	dotted line.  (middle) Radial variation in gas suface density. 
	Filled/open circles represent  molecular gas density along the major 
	axis from Young \et (1995) and Crosthwaite \& Turner (private 
	communication). The squares show the azimuthally-averaged 
	atomic hydrogen surface density (Tilanus \& Allen 1993). (The
 	H densities were multiplied by a factor of 1.4 to include
	the He mass.)  The inner and outer spiral arms imprint the 
	local HI maxima at 120\asec and 240\asec. The dashed line
	is the critical surface density for local gravitational instability.
	(bottom) Ratio of total gas surface density to critical density.
	}  
	\label{fig:m83_gas} \end{figure}
	\begin{figure}[h]
	 \caption{
	Atlas of \Ha + [NII] images and radial surface brightness
	profiles.   A vertical, dotted line marks the radius of the
	threshold in the azimuthally-averaged star formation activity.
	Dashed lines show variations in threshold radius with position
	angle.  North and east are marked with bars 30\asec long.
	}  
	\label{fig:m1} \end{figure}
	\begin{figure}[h]
	 \caption{
	Radial variation of the instability parameter in sub-critical disks.
	If no threshold radius was found, the radius of maximum instability 
	was used to normalize the abscissa.
	}  
	\label{fig:Q_sub} \end{figure}
	\begin{figure}[h] 
	\caption{Gas surface density in sub-critical disks.
	}  
	\label{fig:sd_sub} \end{figure}
	\begin{figure}[h]
	\caption{Mass fraction of atomic gas versus total gas mass at
	the star formation thresholds.  The sub-critical disks are
	represented by open triangles.
	}
	\label{fig:h1h2_mugas}	\end{figure}
	\begin{figure}[h]
	 \caption{Azimuthal variation in the instability parameter.
	The solid lines represent an
	azimuthal average.  The dashed lines are cuts along the major axis.
	Note that in (b)
	the HI surface density was measured
	at position angles of $\pm 90\deg$, and  the molecular gas was measured
	along the major axis which is 60\deg off this HI strip.
	}  
	\label{fig:plots.pa} \end{figure}
	\begin{figure}[h]
	\caption{
	Radial variation in the instability parameter.  The disks
	with the least azimuthal variation in threshold radius are shown.
	The galactocentric radius is normalized to the threshold radius.
	Dotted lines mark the mean ratio of the gas surface density to
	the critical density at the threshold radius.
	}  
	\label{fig:Q_edge} \end{figure}
	\begin{figure}[h]
	\caption{Radial variation in gas surface density.  The disks
	with the least azimuthal variation in threshold radius are shown.
	}  
	\label{fig:sd_edge} \end{figure}
	\begin{figure}[h]
	 \caption{Instability parameter at the star
	formation threshold vs. revised morphological type (RC2 T-type) of the
	galaxy. 
	(Note that galaxy type is an integer quantity, and the decimals
	are used merely to separate points.)
	Filled (open) symbols represent
	surface densities dominated by atomic (molecular) gas, respectively.
	From left to right, the	points represent these galaxies: 
\n4736 (2.0), 
\n4826 (2.1), 
\n4569 (2.3), 
\n4394 (3.0), 
\n4402 (3.1), 
\n4501 (3.3),
\n7331 (3.4), 
\n4579 (3.5), 
\n4689 (4.0), 
\n2903 (4.2), 
\n5055 (4.3),
\n4321 (4.4), 
\n4651 (5.0), 
\n5236 (5.1), 
\n628  (5.2), 
\n4254 (5.3), 
\n4535 (5.4), 
\n4647 (5.5), 
\n2403 (6.0), 
\n6946 (6.1), 
\n4654 (6.3), 
\n925 (7.0), 
\n4713 (7.2), and 
\n4178 (8.0). 
	}  
	\label{fig:alpha} \end{figure}
	\begin{figure}[h]
	 \caption{
	Instability parameter normalized to the location of the
	threshold in the inner disk.  The dotted lines shown the median
	value of the instability parameter at the outer thresholds for
	reference.
	}  
	\label{fig:Q_in} \end{figure}
	\begin{figure}[h]
	\caption{
	Comparison of the Toomre~Q and shear stability criteria
	in the inner disk. The observed
	gas surface density is shown by the solid line for comparison.
	The radial coordinate is normalized to the star formation 
	threshold.
	}  
	\label{fig:shear_n2403} \end{figure}
	\begin{figure}[h]
	\caption{
	Comparison of the Toomre and shear instability parameters at
	the outer thresholds.  The text describes why these six galaxies
	are shown.
	}  
	\label{fig:Q_shear} \end{figure}


\begin{references}

\reference{} Adler, D. S. \& Westpfahl, D. J. 1996, \aj, 111, 735.

\reference{} Bloemen, J. B. G. M. \et 1986, \aa, 154, 25.

\reference{} Bosma, A. 1977, \aa, 57, 375.  

\reference{} Bosma, A. 1981, \aj, 86, 1791.  

\reference{} Bosma, A., Goss, W. M., \& Allen, R. J. 1981, \aa, 93, 106.   

\reference{} Braun, R., Walterbos, R. A. M., Kennicutt, R. C., \& Tacconi, 
L. J. 1994, \apj, 420, 558.

\reference{} Carignan, C., Charbonneau, P., Boulanger, F., \& Viallefond, F. 1990, \aa, 234, 43.

\reference{} Cayatte, V., van Gorkom, J. H., Balkowski, C., \& Kotanyi, C. 1990, \aj, 100, 604. 

\reference{} Dalcanton, J. J., Spergel, D. N., \& Summers, F. J. 1997, \apj, 482, 659.


\reference{} Dickey \et 1990, \apj, 352, 522.

\reference{} Elmegreen, B. 1993, in Star Formation, Galaxies, and the 
Interstellar Medium, ed. J. Franco, F. Ferrini, \& G. Tenorio-Tagle, Cambridge:
Cambridge University Press, p. 337.

\reference{} Elmegreen, B. 1994, \apj, 433, 39.

\reference{} Elmegreen, B. 1995 \mn, 275, 944.

\reference{} Elmegreen, B. 1991, \apj, 378, 139.

\reference{} Elmegreen, B. 1987, \apj, 312, 626. 

\reference{} Elmegreen, B. \& Parravano, A. 1994, \apj, 435, 121. 

\reference{} Federman, S. R., Glassgold, A. E., \& Kwan, J. 1979, \apj, 227, 466.  

\reference{} Ferrarese, L. \et 1996, \apj, 464, 568. 


\reference{} Ferguson, A., Wyse, R. F. G., Gallagher, J. S., \& Hunter, D. A.
 1998, \apj, 506, L19.  

\reference{} Freedman, W. L. \et 1994a, Nature, 371, 757. 

\reference{} Freedman, W. L. \et 1994b, \apj, 427, 628. 

\reference{} Gerin, M., Casoli, F., Combes, F. 1991, \aa, 251, 32.

\reference{} Gu, Q.-S. \et 1996, \aa, 314, 18.

\reference{} Hodge, P., W. \& Kennicutt, R. C. 1983, \apj, 267, 563.

\reference{} Hunter, D., Elmegreen, B. G., \& Baker, A. L. 1998, \apj, 493, 595.   

\reference{} Hunter, D. A. \& Plummer, J. D. 1996, \apj, 462, 732.


\reference{} Gallagher, J.  \& Hunter, D.  1984, \annrev, 22.  


\reference{} Goldreich, P. \& Lynden-Bell, D. 1965, \mn, 130, 97.


\reference{} Graham, J. A. \et 1999, \apj, 516, 626.

\reference{} Guhathakurta, P., Van Gorkom, J. H., Kotanyi, C. G., \& 
Balkowski, C. 1988, 96, 851.  


\reference{} Hughes, S. M. G. \et 1998, \apj, 501, 32. 

\reference{} Jog \& Solomon 1984 \apj 276, 114 and \apj 276, 127.

\reference{} Jogee, S. 1999, Ph. D. Thesis, Yale Univeristy.

\reference{} Jogee, S., Kenney, J. D. P., \& Scoville, N. 2001, \apj, submitted.

\reference{} Kamphuis \& Briggs 1992, \aa, 253, 335. 

\reference{} Kelson, D. D. \et 1996, \apj, 463, 26.  

\reference{} Kenney, J. D. \& Young, J. S. 1989, \apj, 344, 171. 

\reference{} Kenney, J. D. \& Young, J. S. 1988, \apjs, 66, 261.  

\reference{} Kenney, J. D. P, Scoville, N. Z., \& Wilson, C. D. 1991, \apj, 366, 432.  

\reference{} Kennicutt, R. C., \& Kent, S. M. 1983, \aj, 88, 1094.

\reference{} Kennicutt, R. C. 1989, \apj, 344, 685. 

\reference{} Kennicutt, R. C. 1998, \apj, 508, 491.             

\reference{} Kennicutt, R. C., Tamblyn, P., \& Congdon, C. E. 1994, \apj, 435, 22.  

\reference{} Kormendy \& Norman 1979, \apj, 233, 539.

\reference{} Kurucz, R. L. 1992, private communication.

\reference{} Lelievre, M. \& Roy, J.-R. 2000, \apj, 120, 1306. 

\reference{} Lord, S. D. \& Young, J. S. Y. 1990, \apj, 356, 135.   


\reference{} Macri \et 1999, \apj, 521, 155.

\reference{} Mao, S., Mo, H. J., \& White, S.D.M. 1998, \mn, 297, 71.

\reference{} McCall, M. L., Rybski, P. M., \& Shields, G. A. 1985, \apjs, 
57, 1.


\reference{} Miller, J. S. \& Mathews, W. G. 1972, \apj, 172, 593.


\reference{} Mo, H. J., Mao, S., White, S.D.M. 1998, \mn, 295, 319.

\reference{} Newton, K. 1980, \mn, 689. 

\reference{} Nilson, P. 1973, Uppsala General Catalogue of Galaxies, Series V: A vol. 1.

\reference{} Phookun, B. Vogel, S. N. \& Mundy , L. R. 1993, \apj, 418, 113.

\reference{} Pisano, D. J., Wilcots, E. M, \& Elmegreen, B. G. 1998, \apj, 115, 975.

\reference{} Quirk, W. J. 1972, \apj, 176, L9.

\reference{} Rand, R. J. 1993, \apj, 410, 68.

\reference{} Rogstad \et 1974, \apj, 193, 309.

\reference{} Rots, A. H. 1975, \aa, 45, 43. 

\reference{} Rubin, V. C., Waterman, A. H., \& Kenney, D. P. 1999, \aj, 118, 236.

\reference{} Rudnick, G., Rix, H.-W., \& Kennicutt, R. C. 2000, \apj, 538, 569. 

\reference{} Sage, L. J. \& Westpfahl, D. J. 1991, \aa, 242, 371.

\reference{} Sancisi, R., Fraternali, F., Oosterloo, T., \& Moorsel, G. 2000,
in Gas \& Galaxy Evolution, ASP Conference Series, Vol ?, ed. J. E. Hibbard,
M. P. Rupen, and J. H. van Gorkom.

\reference{} Saha, A. \et 1995, \apj, 438, 8.

\reference{} Schaap, W. E., Sancisi, R., \& Swaters, R. A. 2000, \aa, 356, L49.

\reference{} Schaller, G., Schaerer, D., Meynet, G., \& Maeder, A. 1993, \aasup,
96, 269.

\reference{} Schmidt, M. 1959, \apj, 129, 243. 

\reference{} Sellwood, J. A., \& Balbus, S. A. 1999, \apj, 511, 660.

\reference{} Sharina, M. E., Karachentsev, Tikhonov, N. A. 1997, AstL, 23, 373.

\reference{} Shostak, G. S., \& van der Kruit, P. C. 1984, \aa, 132, 20. 

\reference{} Silbermann, N. A. \et 1996, \apj, 470, 1. 

\reference{} Silk, J. 1997 \apj, 481, 703.

\reference{} Skillman, E. D. 1987, in Star Formation in Galaxies, ed. C. J.
Lonsdale Persson (NASA Conf. Pub. CP-2466), p. 263.

\reference{} Sofue, Y. 1996, \apj, 458, 120.

\reference{} Spitzer, 1968, Diffuse Matter In Space (New York: Interscience Publishers).

\reference{} Tacconi, L. J. \& Young, J. S. 1986, \apj, 308, 600. 

\reference{} Tacconi, L. J. \& Young, J. S. 1989, \apjs, 71, 455.  

\reference{} Tan, J. C., Silk, J., \& Balland, C. 1999, \apj, 522, 579.

\reference{} Thornley, M. \et 1999, Ap\& SS, 269, 391.

\reference{} Tilanus, R. P. J. \& Allen, R. J. 1991, \aa, 244, 8. 

\reference{}Tilanus, R.P.J, \& Allen, R. J. 1993, \aa, 274, 707.  

\reference{} Thon, R., Meusinger, H. 1998, \aa, 338, 413.

\reference{} Thornley, M. D. \& Wilson, C. D. 1995, \apj, 447, 616.

\reference{} Toomre, A. 1964 \apj, 139, 1217. 

\reference{} van der Hulst, J. M., \et 1993, \aj, 106, 548.

\reference{} van der Hulst, J. M., Kennicutt, R. C., Crane, P. C.,
\& Rots, A. H. 1988, \aa, 195, 38.  

\reference{} van Zee, L., Haynes, M. P., Salzer, J. J., \& Broeils, A. H. 1996,
\aj, 112, 129.   

\reference{} van Zee, L., Haynes, M. P., Salzer, J. J., \& Broeils, A. H. 1997,
\aj, 113, 1618.

\reference{} van Zee, L., Skillman, E. D., \& Salzer, J. J. 1998, \aj, 116,
1186.  

\reference{} van der Kruit, P. C., Searle, L. 1981a, \aa, 95, 105.

\reference{} van der Kruit, P. C., Searle, L. 1981b, \aa, 95, 116.

\reference{} van der Kruit, P. C., Searle, L. 1982a, \aa, 110, 61.

\reference{} van der Kruit, P. C., Searle, L. 1982b, \aa, 110, 79.

\reference{} van der Kruit, P. C. \& Shostak, G.S. 1984, \aa, 134, 258.

\reference{} Wang \& Silk 1994 \apj, 427, 759.

\reference{} Warmels, R. H. 1988a, \aasup, 72, 19. 

\reference{} Warmels, R. H. 1998b, \aasup, 72, 57. 

\reference{} Warmels, R. H. 1988c, \aasup, 72, 427. 

\reference{} Westpfal, D. J. 1998, \apjs, 115, 203. 

\reference{} Webster, B. L. \& Smith, M. G. 1983, \mn, 204, 743.

\reference{} Wevers, B. M. H. R., van der Kruit, P. C., \& Allen, R. J. 1986, \aasup, 66, 505.

\reference{} Wilson, C. 1995, \apj, 448, 97.

\reference{} Wong, T. \& Blitz, L. 2000, \apj, 540, 771.

\reference{} Young, J. S. \et 1995, \apjs, 98, 219.  

\reference{} Zaritsky, D., Kennicutt, R. C., Huchra, J. 1994, \apj, 420, 87.

\end{references}
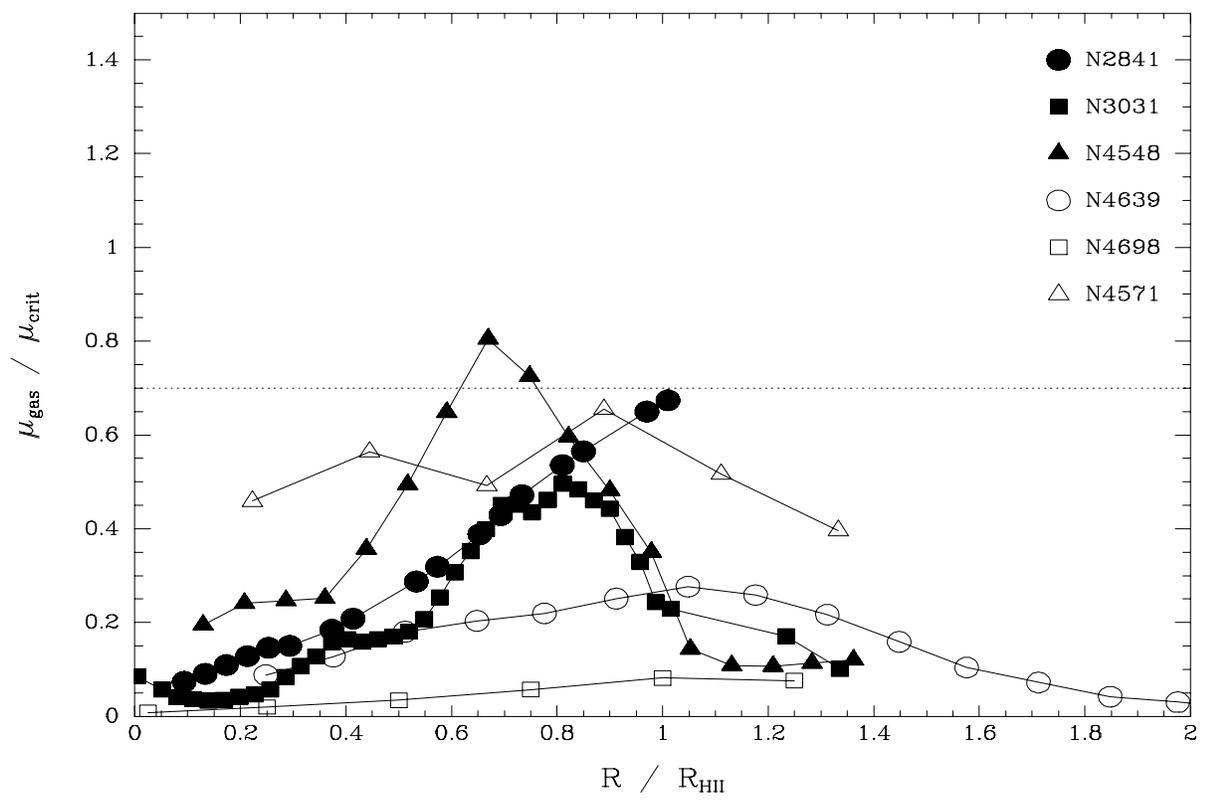
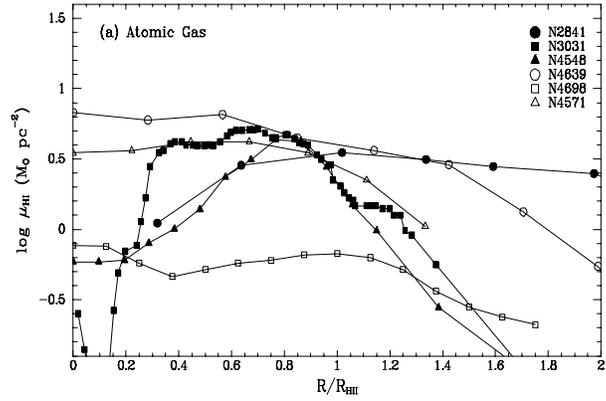
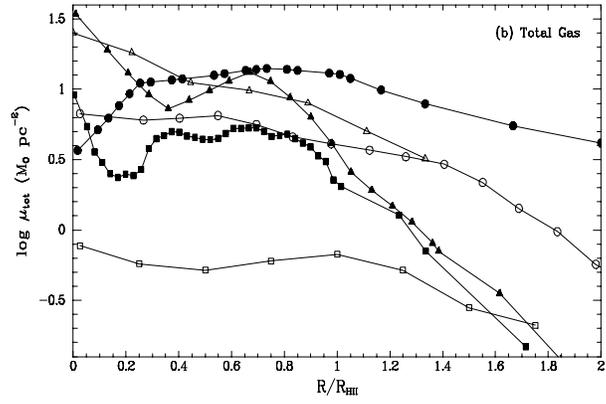
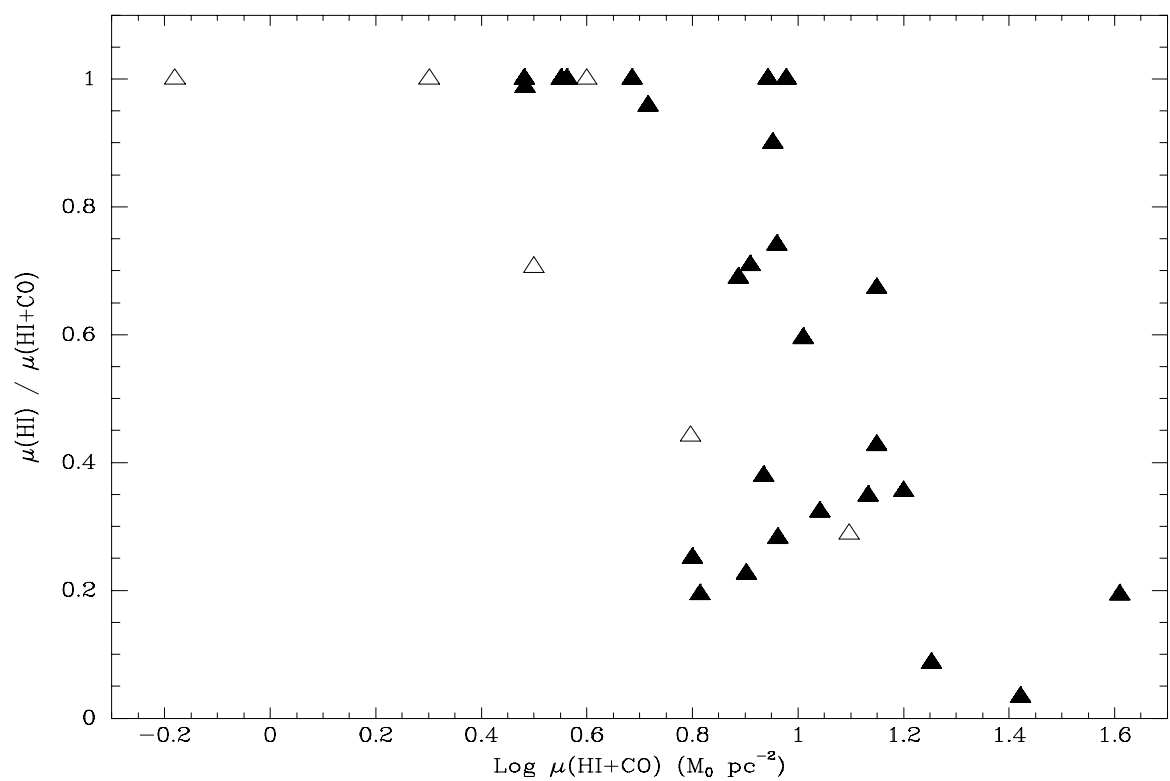
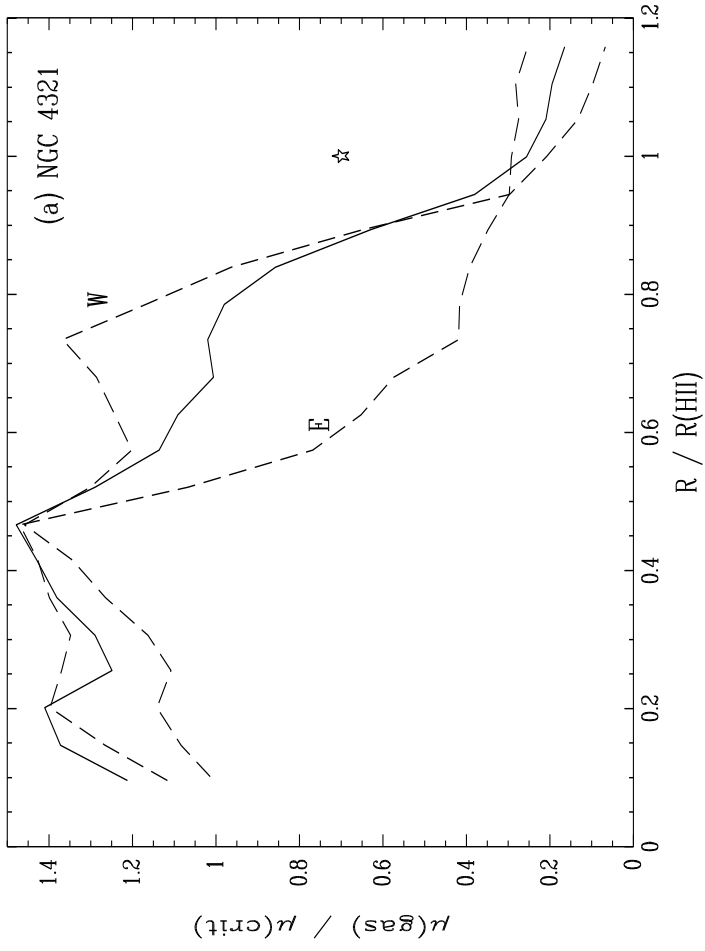
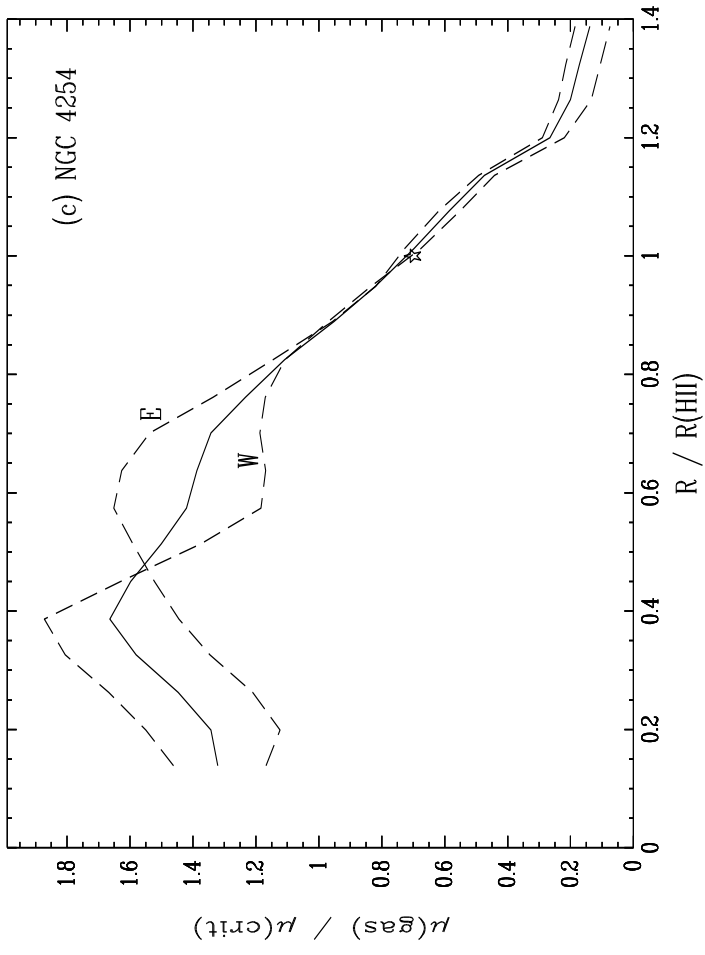
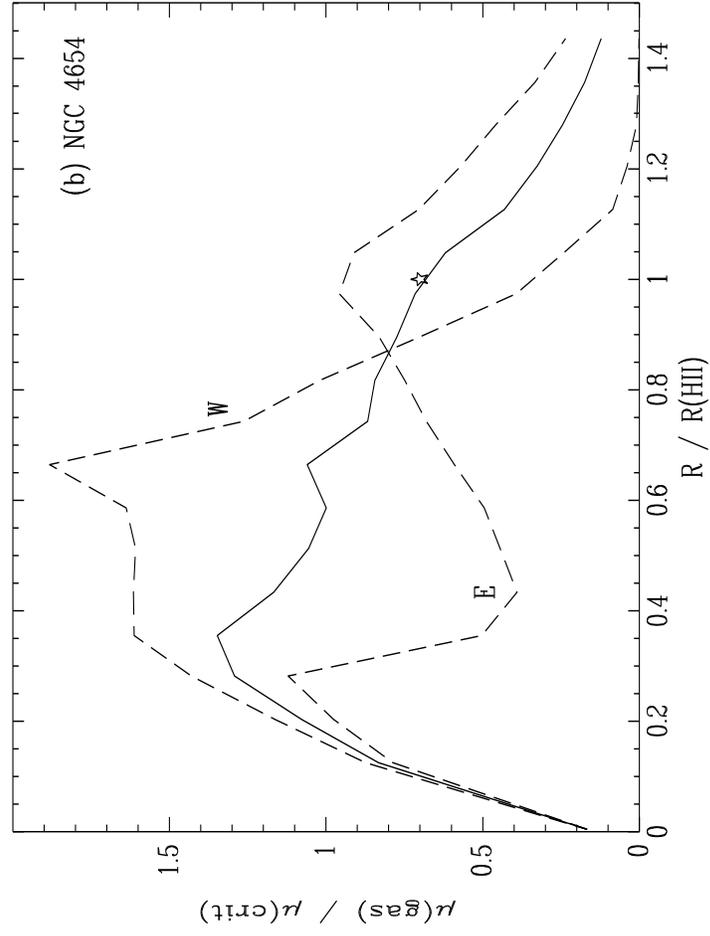
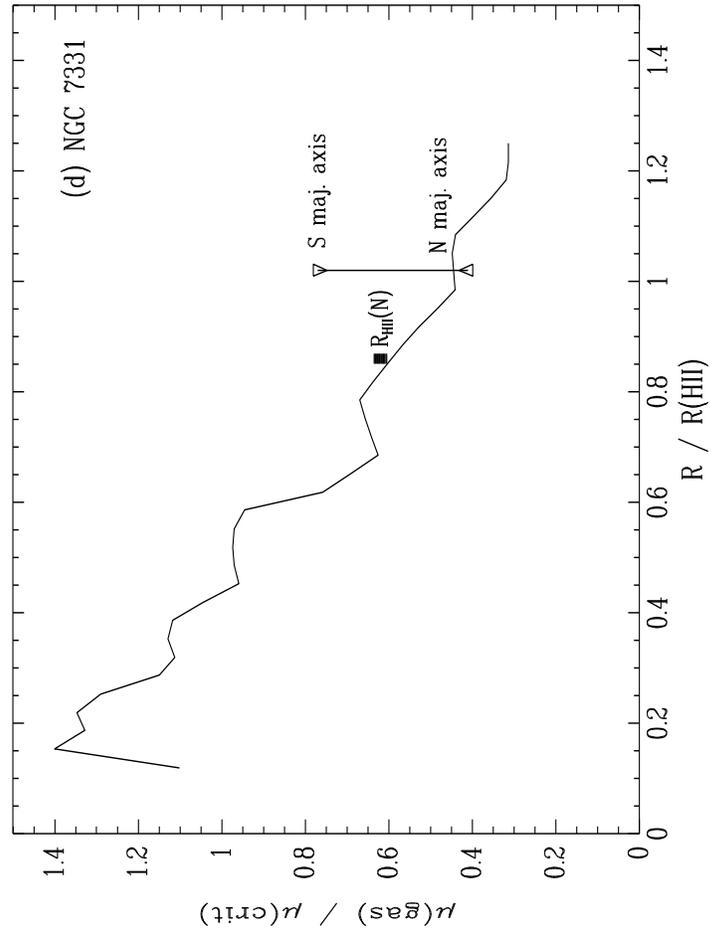
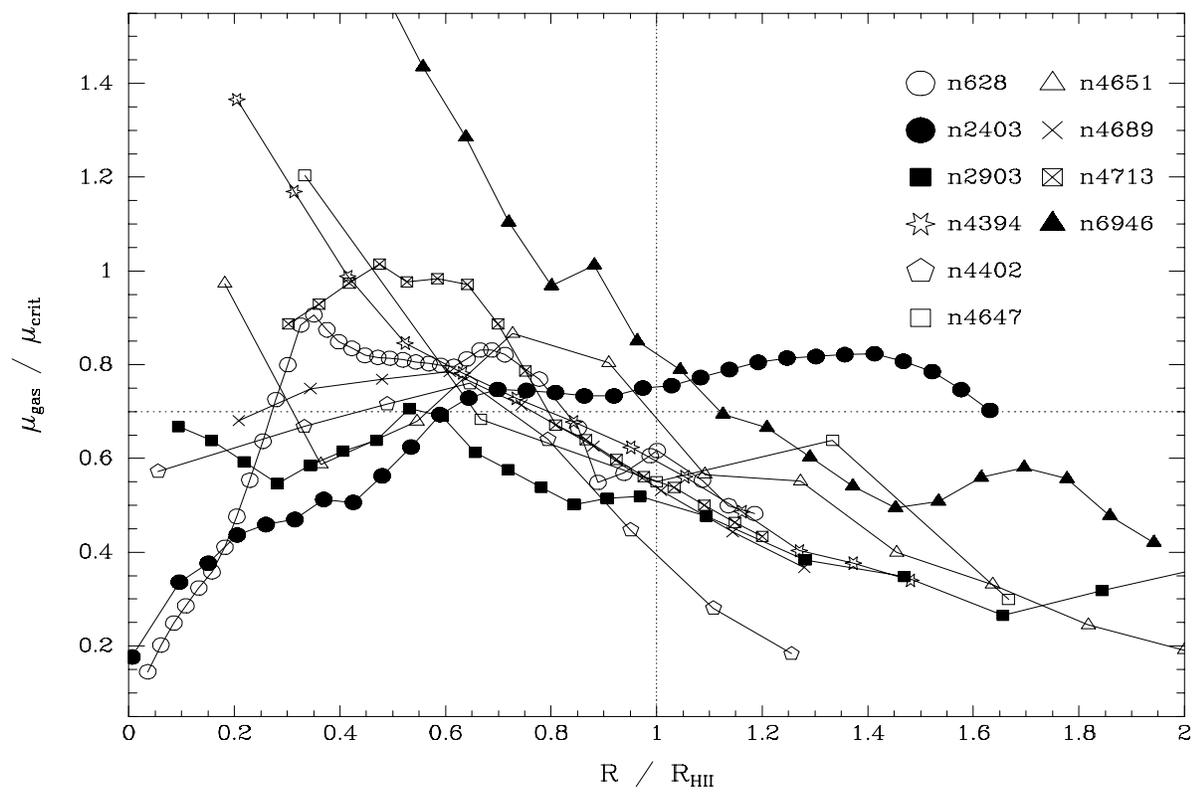
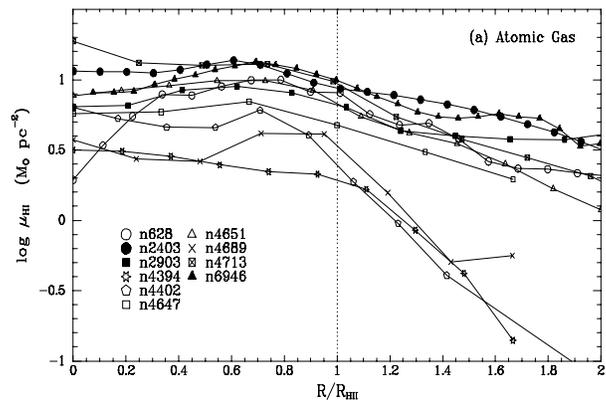
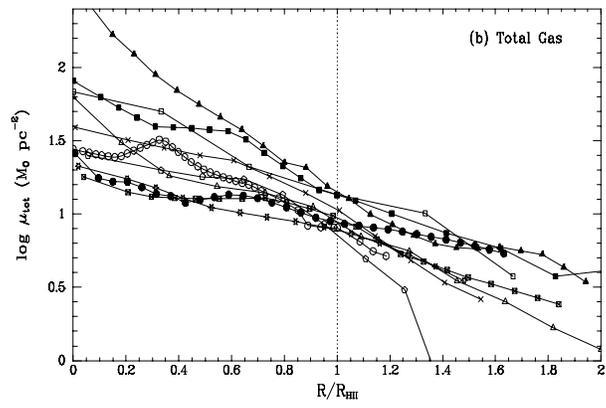
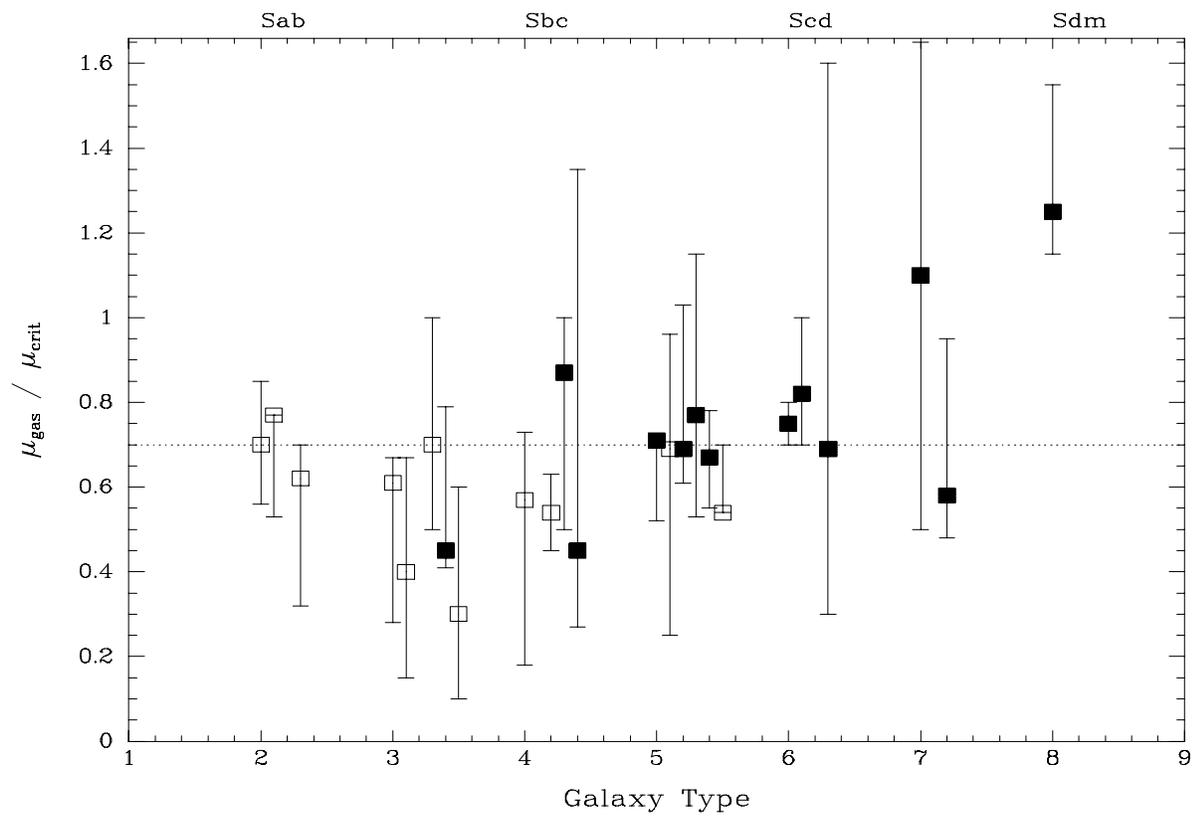
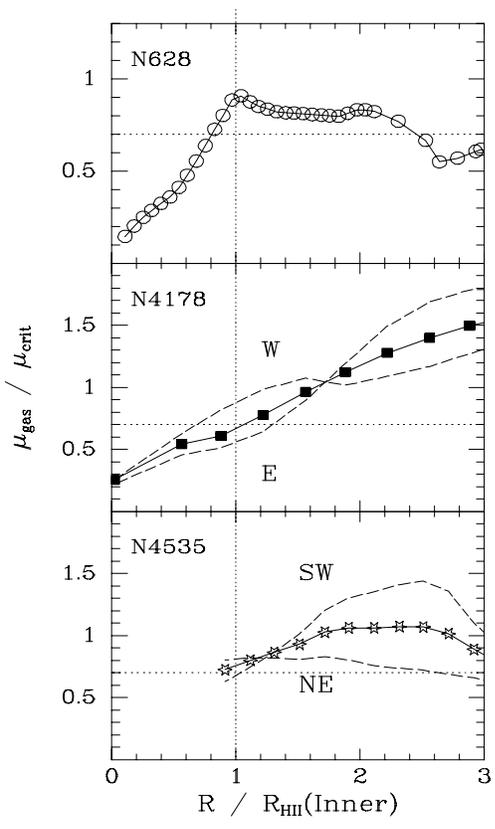
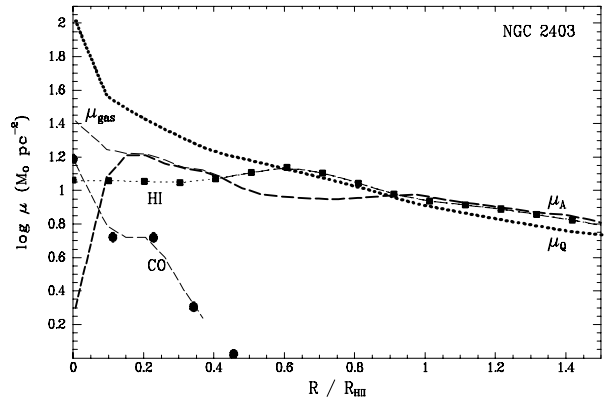
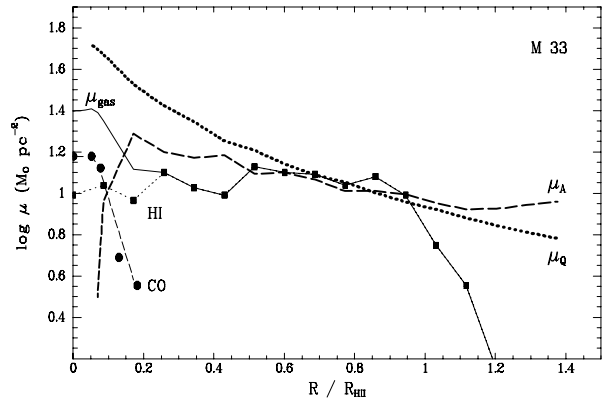
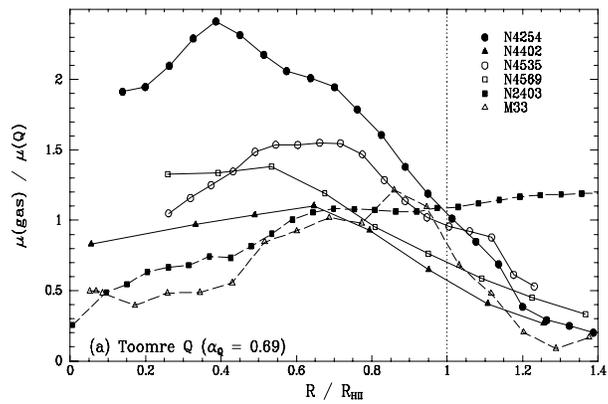
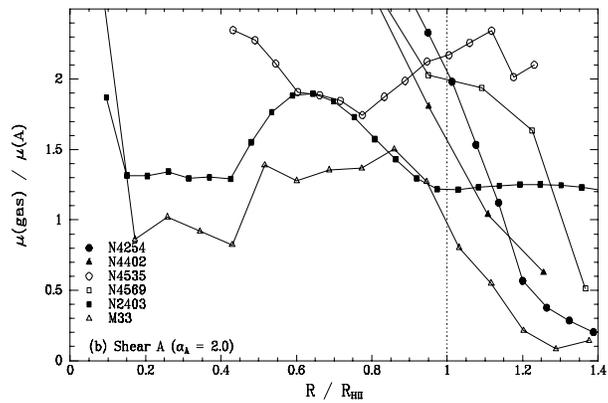
\end{document}